\documentclass[12pt]{article}
\usepackage{axodraw}
\input epsf.tex
\begin{document}
\newcommand{\beq}{\begin{equation}}
\newcommand{\eeq}{\end{equation}}
\newcommand{\beqa}{\begin{eqnarray}}
\newcommand{\eeqa}{\end{eqnarray}}
\newcommand{\beqar}{\begin{eqnarray*}}
\newcommand{\eeqar}{\end{eqnarray*}}
\newcommand{\al}{\alpha}
\newcommand{\be}{\beta}
\newcommand{\del}{\delta}
\newcommand{\D}{\Delta}
\newcommand{\eps}{\epsilon}
\newcommand{\ga}{\gamma}
\newcommand{\Ga}{\Gamma}
\newcommand{\ka}{\kappa}
\newcommand{\inn}{\!\cdot\!}
\newcommand{\h}{\eta}
\newcommand{\kk}{\varphi}
\newcommand\F{{}_3F_2}
\newcommand{\la}{\lambda}
\newcommand{\La}{\Lambda}
\newcommand{\na}{\nabla}
\newcommand{\Om}{\Omega}
\newcommand{\p}{\phi}
\newcommand{\sig}{\sigma}
\renewcommand{\t}{\theta}
\newcommand{\z}{\zeta}
\newcommand{\ssc}{\scriptscriptstyle}
\newcommand{\eg}{{\it e.g.,}\ }
\newcommand{\ie}{{\it i.e.,}\ }
\newcommand{\labell}[1]{\label{#1}} %{\label{#1}} %
\newcommand{\reef}[1]{(\ref{#1})}
\newcommand\prt{\partial}
\newcommand\veps{\varepsilon}
\newcommand\ls{\ell_s}
\newcommand\cF{{\cal F}}
\newcommand\cA{{\cal A}}
\newcommand\cS{{\cal S}}
\newcommand\cT{{\cal T}}
\newcommand\cC{{\cal C}}
\newcommand\cL{{\cal L}}
\newcommand\cG{{\cal G}}
\newcommand\cI{{\cal I}}
\newcommand\cl{{\iota}}
\newcommand\cP{{\cal P}}
\newcommand\cV{{\cal V}}
\newcommand\cg{{\it g}}
\newcommand\cR{{\cal R}}
\newcommand\cB{{\cal B}}
\newcommand\cO{{\cal O}}
\newcommand\tcO{{\tilde {{\cal O}}}}
\newcommand\bz{\bar{z}}
\newcommand\bw{\bar{w}}
\newcommand\hF{\hat{F}}
\newcommand\hA{\hat{A}}
\newcommand\hT{\hat{T}}
\newcommand\htau{\hat{\tau}}
\newcommand\hD{\hat{D}}
\newcommand\hf{\hat{f}}
\newcommand\hg{\hat{g}}
\newcommand\hp{\hat{\phi}}
\newcommand\hi{\hat{i}}
\newcommand\ha{\hat{a}}
\newcommand\hQ{\hat{Q}}
\newcommand\hP{\hat{\Phi}}
\newcommand\hS{\hat{S}}
\newcommand\hX{\hat{X}}
\newcommand\tL{\tilde{\cal L}}
\newcommand\hL{\hat{\cal L}}
\newcommand\tG{{\widetilde G}}
\newcommand\tg{{\widetilde g}}
\newcommand\tphi{{\widetilde \phi}}
\newcommand\tPhi{{\widetilde \Phi}}
\newcommand\te{{\tilde e}}
\newcommand\tk{{\tilde k}}
\newcommand\tf{{\tilde f}}
\newcommand\tF{{\widetilde F}}
\newcommand\tK{{\widetilde K}}
\newcommand\tE{{\widetilde E}}
\newcommand\tpsi{{\tilde \psi}}
\newcommand\tX{{\widetilde X}}
\newcommand\tD{{\widetilde D}}
\newcommand\tO{{\widetilde O}}
\newcommand\tS{{\tilde S}}
\newcommand\tB{{\widetilde B}}
\newcommand\tA{{\widetilde A}}
\newcommand\tT{{\widetilde T}}
\newcommand\tC{{\widetilde C}}
\newcommand\tV{{\widetilde V}}
\newcommand\thF{{\widetilde {\hat {F}}}}
\newcommand\Tr{{\rm Tr}}
\newcommand\tr{{\rm tr}}
\newcommand\STr{{\rm STr}}
\newcommand\M[2]{M^{#1}{}_{#2}}
\parskip 0.3cm
%\begin{document}

%\thispagestyle{empty} \rightline{\small  \hfill IPM/P-2006/xxx}
\vspace*{1cm}

\begin{center}
{\bf \Large  S-matrix elements from  T-duality    }

\vspace*{1cm}

{Komeil Babaei Velni\footnote{komeilvelni@gmail.com} and Mohammad R. Garousi\footnote{garousi@ferdowsi.um.ac.ir} }\\
\vspace*{1cm}
{ Department of Physics, Ferdowsi University of Mashhad,\\ P.O. Box 1436, Mashhad, Iran}
\\
\vspace{2cm}

\end{center}

\begin{abstract}
\baselineskip=18pt

 Recently it has been  speculated  that  the S-matrix elements   satisfy the Ward identity associated with the T-duality. This indicates that a group of S-matrix elements   is invariant  under the linear T-duality transformations on the external states. If one evaluates one component of such T-dual multiplet, then all other components may be found  by the simple use of  the linear T-duality. The assumption   that fields must be independent of the Killing coordinate, however,  may cause, in some cases,  the T-dual multiplet not to be  gauge invariant. In those cases, the S-matrix elements contain  more than one  T-dual multiplet   which are intertwined by the gauge symmetry. 

In this paper, we apply the T-dual Ward identity on the   S-matrix element of one RR $(p-3)$-form   and two   NSNS states on the world volume of a D$_p$-brane to find its corresponding T-dual multiplet. In the case that the RR potential has two transverse indices, the T-dual multiplet is gauge invariant, however, in the case that it has one transverse index the multiplet  is not gauge invariant.  We find a new T-dual multiplet in this case by imposing the   gauge symmetry.  
   We show that the  multiplets are reproduced by explicit calculation, and their    low energy contact terms   at order $\alpha'^2$ are consistent with the existing couplings in the literature.

\end{abstract}

\setcounter{page}{0}
\setcounter{footnote}{0}
\newpage
%\beqa\frac\eeqa
\section{Introduction } \label{intro}

It is known that the  string theory  is invariant  under T-duality \cite{Kikkawa:1984cp,TB,Giveon:1994fu,Alvarez:1994dn,Becker:2007zj} and  S-duality \cite{Font:1990gx,Sen:1994fa,Rey:1989xj,Sen:1994yi,Schwarz:1993cr,Hull:1994ys,Becker:2007zj}. These symmetries  should be carried by  the scattering amplitudes.  It has been speculated in \cite{Garousi:2011we} that these dualities at  linear order should appear in the  amplitudes through the associated Ward identities. This classifies the tree-level amplitudes into T-dual (S-dual) multiplets. Each multiplet includes the scattering amplitudes which interchange under the linear T-duality (S-duality) transformations.

The T-duality holds order by order in string loop expansion \cite{Becker:2007zj}. Hence,  scattering amplitudes at any loop order should satisfy the T-dual Ward identity. This identity dictates that the amplitudes  should be  covariant under linear T-duality transformations on the external states and should be covariant under the full nonlinear T-duality transformations   on the background fields. On the other hand, the S-duality is nonperturbative in string loop expansion \cite{Becker:2007zj}. The S-dual Ward identity then dictates the scattering amplitudes  should be invariant  under linear S-duality transformations on the external states \cite{Garousi:2011vs,Garousi:2011jh, Garousi:2012gh} and should be invariant under the full nonlinear S-duality transformations   on the background fields after including the loops and the nonperturbative effects \cite{Green:1997tv} - \cite{Garousi:2011fc}.

In the T-duality transformations, there is an  assumption   that the background fields must be  independent of the Killing coordinates  along which the T-duality is applied \cite{TB}.   As a result, the T-dual Ward identity can not  capture the T-dual multiplets that are proportional to the momentum along the killing coordinates.   The gauge invariance associated with the massless closed string states, however, intertwine the separate T-dual multiplets. Therefore, one may find all T-dual multiplets by imposing the   Ward identities associated with the T-duality and the gauge transformations. 

The consistency of the standard Chern-Simons couplings of gravity and R-R fields   \cite{Polchinski:1995mt,Douglas:1995bn} with the linear T-duality has been used in  \cite{Becker:2010ij, Garousi:2010rn} as a guiding principle  to find new couplings involving the antisymmetric B-field.  The T-dual multiplet corresponding to the Chern-Simons couplings at order $\alpha'^2$ has been found in \cite{Becker:2010ij,Garousi:2010rn}. It is not invariant under the B-field and the R-R gauge transformations. This reveals that the extension of the Chern-Simons couplings to the T-duality invariant form involves more than one T-dual multiplets.  It has been shown in  \cite{Garousi:2010rn} that the  multiplets  may be found by imposing  the invariance under these  gauge  transformations. Similar consideration should be applied on the scattering amplitudes corresponding to these couplings.

The disk-level scattering amplitude of one RR $(p-3)$-form   and two   NSNS states has been calculated  in \cite{Garousi:2010bm,Garousi:2011ut,Becker:2011ar}. In this paper we would like to apply the T-dual Ward identity on this amplitude to find its corresponding T-dual multiplet. This multiplet contains  the amplitudes for $(p-1)$-form, $(p+1)$-form, $(p+3)$-form, $(p+5)$-form as well as the original amplitude.  As we shall see, in the case that the RR potential has two transverse indices the T-dual multiplet is invariant under the gauge transformations associated with the massless closed strings. However, in the case that the RR potential has one transverse index, the  T-dual multiplet is not invariant under these gauge transformations.  As a result, the S-matrix in this case involves  two T-dual  multiplets. We will  find  the second multiplet  by imposing the   Ward identity associated with the NSNS gauge transformations on the first multiplet.

The outline of the paper is as follows: We begin in section 2 by reviewing the T-duality transformations and the method for finding the  T-dual completion of a S-matrix element.   In section 3.1, we show that the S-matrix element of one RR $(p-3)$-form with two transverse indices and two NSNS states calculated  in \cite{Garousi:2010bm,Garousi:2011ut,Becker:2011ar} does not satisfy the T-dual Ward identity. Imposing this identity, we will   find the T-dual multiplet whose first component is the amplitude of the R-R $(p-3)$-form. This multiplet satisfies the Ward identities associated with the R-R and the NSNS gauge transformation. In section 3.2, we find the T-dual multiplet whose first component is the amplitude of the RR $(p-3)$-form with one transverse index. This   multiplet    does not  satisfy the Ward identity associated with the NSNS  gauge transformations. In sections 3.3, by imposing this identity we find a new  T-dual multiplet.     In section 3.4, we write the combination of these two multiplets in terms of the field strength of B-field. In section 3.5, we will find the T-dual multiplet whose first component is the scattering amplitude of one RR $(p-3)$-form with one transverse index, one B-field and one gauge boson. This multiplet satisfies  all gauge symmetries. In section 4, we find    the low energy contact terms  of these  multiplets at order $\alpha'^2$  and show that they are consistent  with the existing couplings in the literature.

\section{T-duality}

The full set of nonlinear T-duality transformations for massless RR and NSNS fields have been found in \cite{TB,Meessen:1998qm,Bergshoeff:1995as,Bergshoeff:1996ui,Hassan:1999bv}. The nonlinear T-duality transformations of the  RR field $C$ and the antisymmetric field $B$ are such that  the expression ${\cal C}=e^BC$  transforms linearly under  T-duality \cite{Taylor:1999pr}. When  the T-duality transformation acts along the Killing coordinate $y$,  the massless NSNS fields and  ${\cal C}$  transforms as:
\beqa
e^{2\tphi}=\frac{e^{2\phi}}{G_{yy}}&;& 
\tG_{yy}=\frac{1}{G_{yy}}\nonumber\\
\tG_{\mu y}=\frac{B_{\mu y}}{G_{yy}}&;&
\tG_{\mu\nu}=G_{\mu\nu}-\frac{G_{\mu y}G_{\nu y}-B_{\mu y}B_{\nu y}}{G_{yy}}\nonumber\\
\tB_{\mu y}=\frac{G_{\mu y}}{G_{yy}}&;&
\tB_{\mu\nu}=B_{\mu\nu}-\frac{B_{\mu y}G_{\nu y}-G_{\mu y}B_{\nu y}}{G_{yy}}\nonumber\\
{\cal \tC}^{(n)}_{\mu\cdots \nu y}={\cal C}^{(n-1)}_{\mu\cdots \nu }&;&
{\cal \tC}^{(n)}_{\mu\cdots\nu}={\cal C}^{(n+1)}_{\mu\cdots\nu y}\labell{Cy}
\eeqa
where $\mu,\nu$ denote any coordinate directions other than $y$. In above transformation the metric is given in the string frame. If $y$ is identified on a circle of radius $R$, \ie $y\sim y+2\pi R$, then after T-duality the radius becomes $\tilde{R}=\alpha'/R$. The string coupling is also shifted as $\tilde{g}_s=g_s\sqrt{\alpha'}/R$.

We would like to study the T-dual Ward identity of  scattering amplitudes, so we need the above transformations at the linear order. Assuming that the NSNS  fields are small perturbations around the flat space, \eg $G_{\mu\nu}=\eta_{\mu\nu}+h_{\mu\nu}$, the above  transformations take the following linear form:
\beqa
&&
\tilde{\phi}=\phi-\frac{1}{2}h_{yy},\,\tilde{h}_{yy}=-h_{yy},\, \tilde{h}_{\mu y}=B_{\mu y},\, \tilde{B}_{\mu y}=h_{\mu y},\,\tilde{h}_{\mu\nu}=h_{\mu\nu},\,\tilde{B}_{\mu\nu}=B_{\mu\nu}\nonumber\\
&&{\cal \tC}^{(n)}_{\mu\cdots \nu y}={\cal C}^{(n-1)}_{\mu\cdots \nu },\,\,\,{\cal \tC}^{(n)}_{\mu\cdots\nu}={\cal C}^{(n+1)}_{\mu\cdots\nu y}\labell{linear}
\eeqa
The T-duality transformation of the gauge field on the world volume of D-brane, when it is along the Killing direction, is $\tilde{A}_y=\Phi^y$ where $\Phi^y$ is the transverse scalar. When the gauge field is not along the Killing direction it is invariant under the T-duality.

The strategy to find  the couplings which are invariant under T-duality    is given in \cite{Garousi:2009dj}. This method can be used to find the T-dual multiplet corresponding to a given scattering amplitude. Let us review this method here.  
Suppose we are implementing T-duality along a world volume direction $y$ of a D$_p$-brane. As we have illustrated in the table 1, we first separate the  world-volume indices  along and orthogonal to $y$,     and then apply the  T-duality transformations \reef{linear}.  The  orthogonal indices  are  the complete world-volume indices   of the  T-dual D$_{p-1}$-brane. However,  $y$  in the T-dual theory, which is a normal bundle index, is not  complete. 
%One must then include some other terms in the original theory to have totally completed indices in the T-dual theory. 
On the other hand, the normal bundle indices of  the original theory  are not complete in the T-dual D$_{p-1}$-brane. They do not include the $y$ index.  In a T-dual multiplet, the index  $y$   must be combined  with the incomplete normal bundle indices  to make them complete.  If a scattering amplitude is not invariant under the T-duality,  one should then add  new amplitudes  to it  to have   the complete indices in the T-dual theory. In this way one can find the T-dual multiplet. 

\begin{center}
\begin{picture}(130,115)(0,0)
\SetColor{Black}
\Line(0,10)(0,100)
\Line(60,10)(60,100)
\Line(130,10)(130,100)
\Line(-90,80)(140,80)
\Line(-90,40)(140,40)
\Text(30,95)[]{\small D$_p$-brane}
\Text(95,95)[]{\small D$_{p-1}$-brane}
\Text(-60,62)[]{\small world volume index}
\Text(30,62)[]{$(a,y)$}
\Text(90,62)[]{$a$}
\Text(-60,22)[]{\small transverse index}
\Text(30,22)[]{$i$}
\Text(90,22)[]{$(i,y)$}
 \end{picture}
\end{center}
{\sl Table 1 :{\rm The world volume index in the original D$_p$-brane theory is $(a,y)$ and the transverse index is $i$. In the T-dual theory D$_{p-1}$-brane, the world volume index is $a$ and the transverse index is $(i,y)$. In a T-dual theory this index is complete.}
}

\section{ T-dual multiplets}

The disk-level S-matrix element of one RR $(p-3)$-form and two NSNS states has been calculated in \cite{Craps:1998fn,Garousi:2010bm,Garousi:2011ut,Becker:2011ar}. The amplitude is nonzero when the RR polarization has two, one and zero transverse indices. In this paper we are interested in the first two cases.  In the next section, as an easy exercise to apply the strategy for finding the T-dual multiplet, we will find the T-dual multiplet corresponding to the  case that the RR $(p-3)$-form has two transverse indices and in the section 3.2 we will find the T-dual multiplet for the case that the RR potential has one transverse index.

\subsection{RR potential with two transverse indices}

When the RR potential has two transverse indices  the amplitude  is nonzero only for the case that the NSNS polarizations are antisymmetric \cite{ Garousi:2010bm,Becker:2011ar}. The amplitude  is given by the following expression \cite{Garousi:2010bm,Becker:2011ar}:\footnote{Our conventions set $\alpha'=2$ in the string theory amplitudes. Our index convention is that the Greek letters  $(\mu,\nu,\cdots)$ are  the indices of the space-time coordinates, the Latin letters $(a,d,c,\cdots)$ are the world-volume indices and the letters $(i,j,k,\cdots)$ are the normal bundle indices.}
\beqa
{ A}_2(C^{(p-3)})\sim T_p(\veps_1^{(p-3)})_{ij}{}^{a_6\cdots a_p}\eps_{a_0\cdots a_p}p_2^{a_0}p_3^{a_1}p_2^ip_3^j(\veps_2^A)^{a_2a_3}(\veps_3^A)^{a_4a_5}\cI_1\delta^{p+1}(p_1^a+p_2^a+p_3^a) \labell{amp5}
\eeqa
where the function $\cI_1$ has the  following integral representation:
\beqa
\cI_1&=&\int_0^1dr_2\int_0^1dr_3\int_0^{2\pi}d\theta\,\frac{ \tK}{r_2r_3} \labell{I1}
\eeqa
where $\tK$ is 
\beqa
\tK&\!\!\!\!=\!\!\!\!&\ {r_2}^{2p_1\cdot p_2}\ {r_3}^{2 p_1\cdot p_3} (1-{r_2}^2)^{p_2\cdot D\cdot p_2}(1-{r_3}^2)^{p_3\cdot D\cdot p_3}\nonumber\\
&&\left.\times |r_2-r_3 e^{i \t}|^{2 p_2\cdot p_3}|1-r_2 r_3 e^{i \t}|^{2 p_2\cdot D\cdot p_3}\right.  \nonumber
\eeqa
This function  is symmetric under exchanging the momentum labels 2, 3. The subscribe 2 in ${ A}_2$ refers to the number of transverse indices of  the RR potential. In the amplitude \reef{amp5}, $\veps_1$ is the polarization of the RR $(p-3)$-form, $\veps_2^A,\, \veps_3^A$ are the polarizations of the B-fields and $T_p$ is the tension of  D$_p$-brane in the string frame, \ie
 \beqa
 T_p&=&\frac{1}{g_s(2\pi)^p(\alpha')^{(p+1)/2}}\labell{tension}
 \eeqa
 where $g_s=e^{\phi_0}$ is the  closed string coupling, and $\phi_0$ is the constant background dilaton field. 
 
 The amplitude \reef{amp5} satisfies the Ward identity associated with the NSNS gauge transformations. However, it does not satisfy the Ward identity associated with the T-duality and the RR gauge transformations. The T-dual Ward identity in which we are interested in this paper, can be satisfied when the appropriate amplitudes of the RR $(p-1)$-form,   $(p+1)$-form , $(p+3)$-form and  $(p+5)$-form are included. On the other hand, the Ward identity associated with the RR gauge transformations can be satisfied when the amplitude of the RR  $(p-3)$-form with one transverse index is included. 

To study the T-dual Ward identity of the above amplitude,  one should first apply the nonlinear T-duality on the background fields, and then apply the linear T-duality transformations on the external  states. If one implements T-duality along a world volume direction  of the D$_p$-brane, then  the nonlinear T-duality of the background fields leaves invariant the Mandelstam variables, hence, the function $\cI_1$  remains  invariant. The T-duality also  transforms 
\beqa
%e^{-\phi}\sqrt{-\eta}&\longrightarrow&e^{-\phi}\sqrt{-\eta}\nonumber\\
T_p\delta^{p+1}(p_1^a+p_2^a+p_3^a)&\longrightarrow&T_{p-1}\delta^{p}(p_1^a+p_2^a+p_3^a)\labell{backT}
 \eeqa
where  we have used  the assumption implicit in the T-duality that fields are independent of the Killing coordinate, \eg $\delta^{p+1}(p_1^a+p_2^a+p_3^a)=\delta^{p}(p_1^a+p_2^a+p_3^a)\delta(0)$ where $\delta(0)=2\pi R$ and $R$ is the radius of the compact direction on which the T-duality is applied. So the D$_p$-brane  of type IIA/IIB transforms to the D$_{p-1}$-brane  of type IIB/IIA. 

We now apply the linear T-duality on the external states.   From the contraction with the world volume form, one of the indices $a_6,\cdots, a_{p}$ of the RR potential  or the indices  $a_0,\cdots, a_5$ of the B-field polarizations  must include $y$. So there are two cases to consider: First when the RR potential ${\veps_1}^{(p-3)}$ carries the $y$-index and second when   the NSNS states carries the $y$-index. In the former case, one can easily verify that the amplitude is  covariant under the linear T-duality, \ie the amplitude is invariant up to \reef{backT}. In the latter case, however, the amplitude is not covariant because when the B-field carries the $y$-index it transforms to the metric under the linear T-duality \reef{linear}. 

To find the T-dual completion of the S-matrix element \reef{amp5}, we  first  write the factor in \reef{amp5} which contains the B-field polarizations and the momenta along the brane,  as tensor  $k^{a_0\cdots a_5}\equiv p_2^{a_0}p_3^{a_1} (\veps_2^A)^{a_2a_3}(\veps_3^A)^{a_4a_5}$. Then one finds the following six possibilities for $k^{a_0\cdots a_5}$ to carry the Killing index $y$, \ie  
\beqa
\eps_{a_0\cdots a_5a_6\cdots a_p}k^{a_0\cdots a_5}&=&\eps_{a_0\cdots a_4ya_6\cdots a_p}\bigg(-k^{ya_0a_1a_2a_3a_4}+k^{a_0ya_1a_2a_3a_4}-k^{a_0a_1ya_2a_3a_4}\nonumber\\
&&\qquad\qquad\qquad+k^{a_0a_1a_2ya_3a_4}-k^{a_0a_1a_2a_3ya_4}+k^{a_0a_1a_2a_3a_4y}\bigg)\labell{w-v}
\eeqa
Since under the T-duality rules, the external  states must be independent of $y$, any momentum which carries the $y$-index is zero. Apart from the factor $T_p\delta^{p+1}(p_1^a+p_2^a+p_3^a)\cI_1$ whose T-duality transformation has  already been studied, the indices in the amplitude \reef{amp5} can be  separated   as
\beqa
2(\veps_1^{(p-3)})_{ij}{}^{a_6\cdots a_p}\eps_{a_0\cdots a_4ya_6\cdots a_p}p_2^{a_0}p_3^{a_1}p_2^ip_3^j\bigg((\veps_2^A)^{a_3y}(\veps_3^A)^{a_4a_2}+(\veps_2^A)^{a_2a_3}(\veps_3^A)^{a_4y}\bigg) 
\eeqa
Note that because of the presence of the world volume factor $\eps_{a_0\cdots a_4ya_6\cdots a_p}$, the index $y$ in the above expression is a complete index.   Under the linear T-duality transformations \reef{linear} the above expression transforms to the following term on the world volume of D$_{p-1}$-brane:
\beqa
2(\veps_1^{(p-2)})_{ijy}{}^{a_6\cdots a_p}\eps_{a_0\cdots a_4a_6\cdots a_p}p_2^{a_0}p_3^{a_1}p_2^ip_3^j\bigg((\veps_2^S)^{a_3y}(\veps_3^A)^{a_4a_2}+(\veps_2^A)^{a_2a_3}(\veps_3^S)^{a_4y}\bigg) \labell{T}
\eeqa
where $\veps_2^S$ is the polarization of the graviton. The index $y$ which is now a normal bundle index in the    D$_{p-1}$-brane, is not complete any more. This indicates that the original amplitude \reef{amp5} is not consistent with the linear T-duality.

To remedy this failure, one has to add to \reef{amp5}   a new S-matrix element of one RR $(p-1)$-form with three transverse indices and two NSNS states. This amplitude must be such that when it combines with   \reef{amp5}, the indices in the combination  remain complete after  the T-duality transformations. Consider  then the following S-matrix element on the world volume of the D$_p$-brane:
\beqa
{ A}_3(C^{(p-1)})&\sim&2T_p(\veps_1^{(p-1)})_{ijk}{}^{a_5\cdots a_p}\eps_{a_0\cdots a_p}p_2^{a_0}p_3^{a_1}p_2^ip_3^j\bigg((\veps_2^S)^{a_3k}(\veps_3^A)^{a_4a_2}+(2\leftrightarrow 3)\bigg)\nonumber\\
&&\times\cI_1\delta^{p+1}(p_1^a+p_2^a+p_3^a) \labell{amp6}
\eeqa
The combination of the  amplitudes \reef{amp6} and \reef{amp5} may then be covariant under the linear T-duality. In fact  this combination is covariant   only when the world volume Killing coordinate $y$ is carried by the RR potentials. To see this we note that, as we tensioned before, when the RR potential in \reef{amp5} carries  the $y$ index the amplitude is covariant. When the RR potential in \reef{amp6}   carries the $y$ index, it     transforms to $(\veps_1^{(p-2)})_{ijk}{}^{a_6\cdots a_p}$ in which the transverse indices $i,j, k$ are not complete, \ie they do not include $y$. However, the sum of the transformation of the above amplitude and the amplitude \reef{T}  has RR potential $(\veps_1^{(p-2)})_{ijk}{}^{a_6\cdots a_p}$ in which the transverse indices are complete. That is, when  $k=y$  the  amplitude is given by \reef{T}, and  when  $i=y$ or $j=y$ the amplitude is zero because in the amplitude the indices $i,j$ are also the indices of the momenta $p_2^ip_3^j$.

Even though the combination of the amplitudes \reef{amp5} and \reef{amp6} is covariant under the linear T-duality transformations \reef{linear} when the $y$-index is carried by the RR potentials, the amplitude \reef{amp6} is not covariant when the $y$-index is carried by the NSNS polarization tensors.  In the latter case,  apart from the factor $T_p\delta^{p+1}(p_1^a+p_2^a+p_3^a)\cI_1$, the amplitude \reef{amp6} can be written as 
 \beqa
2 (\veps_1^{(p-1)})_{ijk}{}^{a_5\cdots a_p}\eps_{a_0\cdots a_3ya_5\cdots a_p}p_2^{a_0}p_3^{a_1}p_2^ip_3^j\bigg((\veps_2^S)^{yk}(\veps_3^A)^{a_2a_3}+2(\veps_2^S)^{a_2k}
(\veps_3^A)^{a_3y}\bigg)+(2\leftrightarrow 3)\nonumber
\eeqa
where we have used the separation of the indices as in \reef{w-v}. Under the linear T-duality transformations \reef{linear} the above expression transforms to the following terms:
\beqa
2 (\veps_1^{(p)})_{ijky}{}^{a_5\cdots a_p}\eps_{a_0\cdots a_3a_5\cdots a_p}p_2^{a_0}p_3^{a_1}p_2^ip_3^j\bigg((\veps_2^A)^{ky}(\veps_3^A)^{a_2a_3}+2(\veps_2^S)^{a_2k}
(\veps_3^S)^{a_3y}%-2(\veps_2^S)^{a_2y}(\veps_3^S)^{a_3k}+(\veps_2^A)^{a_2a_3}(\veps_3^A)^{ky}
\bigg)+(2\leftrightarrow 3) \labell{non}
\eeqa
The index $y$ which is a normal bundle index in the    D$_{p-1}$-brane, is not complete. As a result, the amplitude \reef{amp6} is not yet consistent with the linear T-duality. 

We have to add another  S-matrix element involving the RR $(p+1)$-form with four transverse indices and two NSNS states to restore the T-duality  of the amplitude \reef{amp6}. To this end, 
we add the following  S-matrix element to \reef{amp6}:%amplitude ${\cal A''}$ that when it combine with the amplitude \reef{A'1}, the indices in the combination must remain complete after T-duality.
\beqa
{ A}_4(C^{(p+1)})&\sim& 4T_p (\veps_1^{(p+1)})_{ijkl}{}^{a_4\cdots a_p}\eps_{a_0\cdots a_p}p_2^{a_0}p_3^{a_1}p_2^ip_3^j\bigg(\frac{1}{2!2!}(\veps_2^A)^{kl}(\veps_3^A)^{a_2a_3}+(\veps_2^S)^{a_2k}
(\veps_3^S)^{a_3l}\bigg)\nonumber\\
&&\times\cI_1\delta^{p+1}(p_1^a+p_2^a+p_3^a)+(2\leftrightarrow 3)\labell{A''1}
\eeqa
When the RR potential in the above amplitude carries the $y$-index, the amplitude by itself is not covariant under the linear T-duality, however, the combination of the amplitudes \reef{amp6} and  \reef{A''1} are covariant when the RR potential in \reef{A''1} and the NSNS polarizations in \reef{amp6} carry the index $y$. That is,  the RR potential in \reef{A''1}    transforms to $(\veps_1^{(p)})_{ijkl}{}^{a_5\cdots a_p}$ in which the transverse indices $i,j, k, l$ do not include $y$. The combination of this transformation and the transformation of \reef{amp6} which is given by \reef{non}, however, has complete transverse indices. %The amplitude for $k=y$ or $l=y$  is given by \reef{non}, and  the amplitude for $i=y$ or $j=y$ is zero because the momentum along the $y$-direction is zero.

Again the amplitude \reef{A''1} is not covariant when the $y$-index is carried by the NSNS polarization tensors.  Separating the world volume indices as in \reef{w-v}, apart from the overall  factor, one can write   the amplitude   as 
\beqa
  4  (\veps_1^{(p+1)})_{ijkl}{}^{a_4\cdots a_p}\eps_{a_0a_1a_2 ya_4\cdots a_p}\bigg(\frac{1}{2!}(\veps_2^A)^{kl}(\veps_3^A)^{a_2y}+(\veps_2^S)^{a_2k}
(\veps_3^S)^{yl}+(\veps_2^S)^{yk}
(\veps_3^S)^{a_2l}\bigg) +(2\leftrightarrow 3)\nonumber
\eeqa
which transforms to the following expression under the linear T-duality \reef{linear}:
\beqa
  4  (\veps_1^{(p+2)})_{ijkly}{}^{a_4\cdots a_p}\eps_{a_0a_1a_2 a_4\cdots a_p}\bigg(\frac{1}{2!}(\veps_2^A)^{kl}(\veps_3^S)^{a_2y}+(\veps_2^S)^{a_2k}
(\veps_3^A)^{ly}+(\veps_2^A)^{ky}
(\veps_3^S)^{a_2l}\bigg) +(2\leftrightarrow 3)\nonumber
\eeqa
Inspired by this, one realizes that there should be the following S-matrix element of one RR $(p+3)$-form with five transverse indices and two NSNS states:  
\beqa
{ A}_5(C^{(p+3)})&\sim& 2T_p(\veps_1^{(p+3)})_{ijklm}{}^{a_3\cdots a_p}\eps_{a_0\cdots a_p}p_2^{a_0}p_3^{a_1}p_2^ip_3^j(\veps_2^A)^{kl}(\veps_3^S)^{a_2m}\nonumber\\
&&\times\cI_1\delta^{p+1}(p_1^a+p_2^a+p_3^a)+(2\leftrightarrow 3)\labell{A'''1}
\eeqa
In this case also the T-duality of the above amplitude when the $y$-index is carried by the RR potential, and the T-duality of the amplitude \reef{A''1} when the $y$-index is carried by the NSNS polarizations, produce  terms with complete transverse indices. However, when the $y$-index is carried by the NSNS polarizations in the above amplitude, one again finds that it is not covariant. To restore the symmetry, one has to add the following S-matrix element of one RR $(p+5)$-form with six transverse indices and two NSNS states:
\beqa
{ A}_6(C^{(p+5)})&\sim& \frac{1}{2}T_p(\veps_1^{(p+5)})_{ijklmn}{}^{a_2\cdots a_p}\eps_{a_0\cdots a_p}p_2^{a_0}p_3^{a_1}p_2^ip_3^j(\veps_2^A)^{kl}(\veps_3^A)^{mn}\nonumber\\
&&\times\cI_1\delta^{p+1}(p_1^a+p_2^a+p_3^a)+(2\leftrightarrow 3)\labell{A''''1}
\eeqa
In this case,  no world volume index is carried by the NSNS polarization tensors, hence, there is no more S-matrix element to be added to the above amplitude. 

Hence, T-duality produces the following sequence: 
\beqa
 A_2\rightarrow A_3\rightarrow A_4\rightarrow A_5\rightarrow A_6\nonumber
 \eeqa
where $A_2,\, A_3,\, A_4,\, A_5$ and $A_6$ are given in  \reef{amp5}, \reef{amp6}, \reef{A''1}, \reef{A'''1} and \reef{A''''1}, respectively. The combination of all these terms  satisfies the T-dual Ward identity.  They form the following T-dual multiplet: 
\beqa
A&\sim&T_p\eps_{a_0\cdots a_p}p_2^{a_0}p_3^{a_1}p_2^ip_3^j\bigg[\frac{1}{2}(\veps_1^{(p-3)})_{ij}{}^{a_6\cdots a_p} (\veps_2^A)^{a_2a_3}(\veps_3^A)^{a_4a_5}+2(\veps_1^{(p-1)})_{ijk}{}^{a_5\cdots a_p} (\veps_2^S)^{a_3k}(\veps_3^A)^{a_4a_2}\nonumber\\
&&+4(\veps_1^{(p+1)})_{ijkl}{}^{a_4\cdots a_p} \bigg(\frac{1}{2!2!}(\veps_2^A)^{kl}(\veps_3^A)^{a_2a_3}+(\veps_2^S)^{a_2k}
(\veps_3^S)^{a_3l}\bigg)\labell{mul}\\
&&+2(\veps_1^{(p+3)})_{ijklm}{}^{a_3\cdots a_p} (\veps_2^A)^{kl}(\veps_3^S)^{a_2m}+\frac{1}{2} (\veps_1^{(p+5)})_{ijklmn}{}^{a_2\cdots a_p} (\veps_2^A)^{kl}(\veps_3^A)^{mn}\bigg]\cI_1 +(2\leftrightarrow 3)\nonumber
\eeqa
where we have dropped the delta function. One can easily verify that the above T-dual multiplet satisfies the Ward identity associated with the NSNS gauge transformations. This can be illuminated by witting each amplitude in terms of the B-field strength $H$. On the other hand, the amplitude satisfies the Ward identity associated with the RR gauge transformations when the amplitudes of the RR $(p-3)$-form with one transverse index, the RR $(p-1)$-form with two transverse indies, the RR $(p+1)$-form with three transverse indices, the RR $(p+3)$-form with four transverse indices and the appropriate amplitude of the RR $(p+5)$-form with five transverse indices are included. Such amplitudes appear in the section 3.2.
%,, \eg $p_1^{a_1}(\veps_1^{(p+5)})_{ijklmn}{}^{a_2\cdots a_p}\eps_{a_0\cdots a_p}$ in the last term can be rewritten as $(F^{(p+6)}_1)_{ijklmn}{}^{a_1\cdots a_p}\eps_{a_0\cdots a_p}$ where we have used the fact that we are only considering the RR potential  $C^{(p+5)}$  which carries  six transverse indices. 

 \subsubsection{Explicit calculations}
 
The explicit calculations of the scattering amplitude of one RR n-form in $(-3/2,-1/2)$-picture and two NSNS vertex operators in $(0,0)$-picture on the world volume of D$_p$-brane involves the following trace of gamma matrices \cite{Garousi:2010bm}:
 \beqa
T(n,p,m)
& =&\frac{1}{n!(p+1)!}\veps_{1\nu_1\cdots \nu_{n}}\eps_{a_0\cdots a_p}A_{[\alpha_1\cdots \alpha_m]}\Tr(\gamma^{\nu_1}\cdots \gamma^{\nu_{n}}\gamma^{a_0}\cdots\gamma^{a_p}\gamma^{\alpha_1\cdots \alpha_m})\labell{trace}
 \eeqa
where $A_{[\alpha_1\cdots \alpha_m]}$ is an antisymmetric combination of the momenta and/or the polarizations of the NSNS states and $m=0,2,4,6,8$. The scattering amplitude is non-zero only  for the following sequence \cite{Garousi:2010bm}:
\beqa
n=p-3,\, n=p-1,\, n=p+1,\, n=p+3,\, n=p+5 \labell{seq}
\eeqa
 The amplitude \reef{amp5} corresponds to the first term of this sequence when the RR potential has two transverse indices \cite{Garousi:2010bm}. The trace \reef{trace} in this case is nonzero for $m=8$ . All other components of the T-dual multiplet \reef{mul} should be  correspond to the other terms of the above  sequence for $m=8$, \ie
 \beqa
 A&\stackrel{?}{\sim}&T_p\delta^{p+1}(p_1^a+p_2^a+p_3^a)\sum_{j=0}^{4} T(p-3+2j,p,8)\cI_1\labell{rel}
 \eeqa
 where the   first term  of the summation has two transverse indices, the second term has three transverse indices and so on.   For $m=8$, there is only one  $A_{[\alpha_1\cdots \alpha_8]}$ which is given by \cite{Garousi:2010bm}
\beqa
A_{[\alpha_1\cdots \alpha_8]}&=&(\veps_3\inn D)_{[\alpha_3\alpha_1}(\veps_2\inn D)_{\alpha_7\alpha_5}(ip_2)_{\alpha_8}(ip_2\inn D)_{\alpha_6}(ip_3)_{\alpha_4}(ip_3\inn D)_{\alpha_2]}\labell{Am8}
\eeqa 
 Using the above expression, one can easily verify the relation \reef{rel}. Hence, the T-dual multiplet \reef{mul} predicted by T-duality is reproduced by explicit calculation.

\subsection{RR potential with one transverse index}

In this section, we are going to find the T-dual multiplet whose first component is the RR potential  $C^{(p-3)}$ with one transverse index. The S-matrix elements of one such RR and two NSNS states is 
  nonzero when the NSNS polarizations are antisymmetric \cite{Garousi:2010bm,Becker:2011ar}. The amplitude in the string frame is given by the following expression \cite{Garousi:2010bm}:
\beqa
{\cal A}_1(C^{(p-3)})&\!\!\!\!\!\sim\!\!\!\!\!&(\veps_1^{(p-3)})_i{}^{a_5\cdots a_p}\eps_{a_0\cdots a_p}p_2^{a_2}(\veps_2^A)^{a_1a_4}\bigg[-2(p_1\inn N\inn\veps_3^A)^{a_3}p_2^{i}p_3^{a_0}\cI_1\nonumber\\
&&+(p_2\inn N\inn\veps_3^A)^{a_3}p_2^{i}p_3^{a_0}\cI_2
+(p_2\inn V\inn\veps_3^A)^{a_3}p_3^{i}p_3^{a_0}\cI_2+\frac{1}{2}(p_2\inn V\inn p_3)p_3^{i}(\veps_3^A)^{a_3a_0}\cI_2\nonumber\\
&&-(p_2\inn V\inn\veps_3^A)^{a_3}p_2^{i}p_3^{a_0}\cI_3-(p_2\inn N\inn\veps_3^A)^{a_3}p_3^{i}p_3^{a_0}\cI_3-\frac{1}{2}(p_2\inn N\inn p_3)p_3^{i}(\veps_3^A)^{a_3a_0}\cI_3\nonumber\\
&&+4(p_3\inn V\inn\veps_3^A)^{a_3}p_2^{i}p_3^{a_0}\cI_4+(p_3\inn V\inn p_3)p_2^{i}(\veps_3^A)^{a_3a_0}\cI_4
\bigg]+(2\leftrightarrow 3)\labell{n=1polarization}
\eeqa
where the  matrices  $N_{\mu\nu}$ and $V_{\mu\nu}$  project spacetime vectors into transverse and parallel subspace to the D$_p$-brane, respectively, the function  $\cI_1$ is the  one   appears in \reef{I1} and $\cI_2,\cI_4$ are 
\beqa
%\cI_1&=&-\int_0^1dr_2\int_0^1dr_3\frac{1}{r_2r_3}\int_0^{2\pi}d\theta\tK\nonumber\\
\cI_2&=&2\int_0^1dr_2\int_0^{1}dr_3\frac{(1-r_2^2)}{r_2}\int_0^{2\pi}d\theta\frac{[r_3(1+r_2^2)-r_2(1+r_3^2)\cos(\theta)]\tK}{|1-r_2r_3e^{i\theta}|^2|r_2-r_3e^{-i\theta}|^2}\nonumber\\
%\cI_3-\cI_2&=&2\int_0^1dr_2\int_0^{1}dr_3\frac{1}{r_2r_3}\int_0^{2\pi}d\theta\frac{(r_2^2-r_3^2)\tK}{(r_2-r_3e^{i\theta})(r_2-r_3e^{-i\theta})}\nonumber\\
%\cI_3+\cI_2&=&2\int_0^1dr_2\int_0^{1}dr_3\frac{1}{r_2r_3}\int_0^{2\pi}d\theta\frac{(1-r_2^2r_3^2)\tK}{(1-r_2r_3e^{i\theta})(1-r_2r_3e^{-i\theta})}\nonumber\\
\cI_4&=&-\int_0^1dr_2\int_0^{1}dr_3\frac{(1+r_3^2)}{r_2r_3(1-r_3^2)}\int_0^{2\pi}d\theta\tK\nonumber
%\cI_7&=&-2\pi\int_0^1dr_2\int_0^{1}dr_3\frac{(1+r_2^2)\tK}{r_2r_3(1-r_2^2)}\nonumber
\eeqa
The function $ \cI_3$  is the same as the function $ \cI_2$   in which the momentum labels 2, 3   are exchanged.
These functions  satisfies the following relations \cite{Garousi:2010bm}:
\beqa
-2p_1\inn N\inn p_3\cI_1+2p_3\inn V\inn p_3\cI_4+p_2\inn N\inn p_3\cI_2-p_2\inn V\inn p_3\cI_3&=&0\nonumber\\
-2p_1\inn N\inn p_2\cI_1+2p_2\inn V\inn p_2\cI_7+p_2\inn N\inn p_3\cI_3-p_2\inn V\inn p_3\cI_2&=&0\labell{identity}
\eeqa
where the  function $ \cI_7$  is the same as the function $ \cI_4$   in which the momentum labels 2, 3   are exchanged. In the amplitude \reef{n=1polarization},
 we have dropped the factor $T_p\delta^{p+1}(p_1^a+p_2^a+p_3^a)$ which is covariant under the T-duality transformations. 
 
 The amplitude satisfies the Ward identity associated with the NSNS states \cite{Garousi:2010bm}. However, it satisfies the RR gauge transformation when one includes the amplitude of the RR $(p-3)$-form with two transverse indices \reef{amp5}, and the amplitude of the RR $(p-3)$-form with no transverse index \cite{Garousi:2011ut}.  In fact, using the above identity one finds  the combination of the last terms in the second, the third and the fourth lines of  \reef{n=1polarization} is proportional to  $ \cI_1$ . When they combine with the amplitude \reef{amp5}, the result can be written in terms of  the RR field strength. It has been shown in \cite{Garousi:2011ut} that all other terms in the amplitude \reef{n=1polarization}  can be combined with the appropriate terms in the scattering amplitude of the RR $(p-3)$-form with no transverse index to be written in terms of the RR field strength. 
 
Let us study the T-duality of the amplitude \reef{n=1polarization}. Since the transverse index $i$ of the RR potential is contracted with  momentum, the amplitude is invariant under the linear T-duality when the Killing index $y$ is carried by the RR potential. However, when the $y$-index is carried by the NSNS polarization tensors, one finds that the amplitude is not invariant. Doing the same steps as we have done in the previous section, one  finds that  the following S-matrix element has to be added to the  amplitude \reef{n=1polarization}:
\beqa
{\cal A}_2(C^{(p-1)})&\!\!\!\!\!\sim\!\!\!\!\!&  (\veps_1^{(p-1)})_{ij}{}^{a_4\cdots a_p}\eps_{a_0\cdots a_p}p_2^{a_3}\bigg[
2p_2^i p_3^{a_0}\bigg(2(\veps_2^S)^{a_2j}(p_1\inn N\inn\veps_3^A)^{a_1}+(\veps_2^A)^{a_1a_2}(p_1\inn N\inn\veps_3^S)^j\bigg)\cI_1\nonumber\\
&&-p_2^i p_3^{a_0}\bigg(2(\veps_2^S)^{a_2j}(p_2\inn N\inn \veps_3^A)^{a_1}+(\veps_2^A)^{a_1a_2}(p_2\inn N\inn\veps_3^S)^j\bigg)\cI_2\nonumber\\
&&-p_3^i p_3^{a_0}\bigg(2(\veps_2^S)^{a_2j}(p_2\inn V\inn\veps_3^A)^{a_1}+(\veps_2^A)^{a_1a_2}(p_2\inn V\inn\veps_3^S)^j\bigg)\cI_2\nonumber\\
&&+p_3^i p_2\inn V\inn p_3\bigg((\veps_2^S)^{a_2j}(\veps_3^A)^{a_0a_1}+(2\leftrightarrow 3)\bigg)\cI_2\labell{n=2polarization}\\
&&+p_2^i p_3^{a_0}\bigg(2(\veps_2^S)^{a_2j}(p_2\inn V\inn \veps_3^A)^{a_1}+(\veps_2^A)^{a_1a_2}(p_2\inn V\inn\veps_3^S)^j\bigg)\cI_3\nonumber\\
&&+p_3^i p_3^{a_0}\bigg(2(\veps_2^S)^{a_2j}(p_2\inn N\inn\veps_3^A)^{a_1}+(\veps_2^A)^{a_1a_2}(p_2\inn N\inn\veps_3^S)^j\bigg)\cI_3\nonumber\\
&&-p_3^i p_2\inn N\inn p_3\bigg((\veps_2^S)^{a_2j}(\veps_3^A)^{a_0a_1}+(2\leftrightarrow 3)\bigg)\cI_3\nonumber\\
&&-4p_2^i p_3^{a_0}\bigg(2(\veps_2^S)^{a_2j}(p_3\inn V\inn \veps_3^A)^{a_1}+(\veps_2^A)^{a_1a_2}(p_3\inn V\inn\veps_3^S)^j\bigg)\cI_4\nonumber\\
&&+2p_2^i p_3\inn V\inn p_3\bigg((\veps_2^S)^{a_2j}(\veps_3^A)^{a_0a_1}+(2\leftrightarrow 3)\bigg)\cI_4\bigg]+(2\leftrightarrow 3)\nonumber
\eeqa
The above amplitude satisfies the Ward identity associated with  the antisymmetric NSNS gauge transformations which is inherited from the original amplitude \reef{n=1polarization}. However, the amplitude does not   satisfy the Ward identity associated with the  symmetric NSNS gauge transformations. This indicates that there should be another T-dual multiplet which is not related to the above amplitude by T-duality.   Taking into account that the consistency of the S-matrix elements  with the linear T-duality does not fix the terms which are proportional to the momentum in the $y$ direction, one realizes that the new multiplet  must be proportional to $p_2^j$ or $p_3^j$. %In the Appendix we will find the terms which make the above amplitude to be invariant under the NSNS gauge transformation. 
In the next section we will find such T-dual multiplet by imposing the Ward identity on the  symmetric NSNS state. 

The amplitude \reef{n=2polarization} would satisfy the Ward identity associated with the RR gauge transformations when it combines with the amplitude \reef{amp6} and the  amplitude of the RR $(p-1)$-form with one transverse index in which we are not interested in this paper. In fact, using the identity \reef{identity} one finds the combination of   the terms in the fourth, the seventh and the last lines of \reef{n=2polarization} is proportional to $\cI_1$. As a result, the sum of these terms and the amplitude \reef{amp6} satisfies the RR gauge transformations. One can show that all other terms are proportional to the RR momentum which is a necessary  condition for satisfying the RR gauge transformations. 

 The combination of the amplitude  \reef{n=1polarization} and \reef{n=2polarization} is covariant under the T-duality \reef{linear} when the $y$-index   is carried by the RR potential. In the case that the $y$-index is carried by the NSNS polarizations in \reef{n=2polarization}, the amplitude is not covariant unless one combines it with  the following amplitude:
\beqa
{\cal A}_3(C^{(p+1)})&\!\!\!\!\!\sim\!\!\!\!\!&- 2 (\veps_1^{(p+1)})_{ijk}{}^{a_3\cdots a_p}\eps_{a_0\cdots a_p}p_2^{a_2}\bigg[
2p_2^i p_3^{a_0}\bigg(\frac{1}{2!}(\veps_2^A)^{jk}(p_1\inn N\inn\veps_3^A)^{a_1}-(\veps_2^S)^{a_1j}(p_1\inn N\inn\veps_3^S)^k\bigg)\cI_1\nonumber\\
&&-p_2^i p_3^{a_0}\bigg(\frac{1}{2!}(\veps_2^A)^{jk}(p_2\inn N\inn \veps_3^A)^{a_1}-(\veps_2^S)^{a_1j}(p_2\inn N\inn\veps_3^S)^k\bigg)\cI_2\nonumber\\
&&-p_3^i p_3^{a_0}\bigg(\frac{1}{2!}(\veps_2^A)^{jk}(p_2\inn V\inn\veps_3^A)^{a_1}-(\veps_2^S)^{a_1j}(p_2\inn V\inn\veps_3^S)^k\bigg)\cI_2\nonumber\\
&&+\frac{1}{2}p_3^i p_2\inn V\inn p_3\bigg(\frac{1}{2!}(\veps_2^A)^{jk}(\veps_3^A)^{a_0a_1}+(\veps_3^S)^{a_0j}(\veps_2^S)^{a_1k}+(2\leftrightarrow 3)\bigg)\cI_2\nonumber\\
&&+p_2^i p_3^{a_0}\bigg(\frac{1}{2!}(\veps_2^A)^{jk}(p_2\inn V\inn \veps_3^A)^{a_1}-(\veps_2^S)^{a_1j}(p_2\inn V\inn\veps_3^S)^k\bigg)\cI_3\labell{n=3polarization}\\
&&+p_3^i p_3^{a_0}\bigg(\frac{1}{2!}(\veps_2^A)^{jk}(p_2\inn N\inn\veps_3^A)^{a_1}-(\veps_2^S)^{a_1j}(p_2\inn N\inn\veps_3^S)^k\bigg)\cI_3\nonumber\\
&&-\frac{1}{2}p_3^i p_2\inn N\inn p_3\bigg(\frac{1}{2!}(\veps_2^A)^{jk}(\veps_3^A)^{a_0a_1}+(\veps_3^S)^{a_0j}(\veps_2^S)^{a_1k}+(2\leftrightarrow 3)\bigg)\cI_3\nonumber\\
&&-4p_2^i p_3^{a_0}\bigg(\frac{1}{2!}(\veps_2^A)^{jk}(p_3\inn V\inn \veps_3^A)^{a_1}-(\veps_2^S)^{a_1j}(p_3\inn V\inn\veps_3^S)^k\bigg)\cI_4\nonumber\\
&&+p_2^i p_3\inn V\inn p_3\bigg( \frac{1}{2!}(\veps_2^A)^{jk}(\veps_3^A)^{a_0a_1}+(\veps_3^S)^{a_0j}(\veps_2^S)^{a_1k}+(2\leftrightarrow 3)\bigg)\cI_4\bigg]+(2\leftrightarrow 3)\nonumber
\eeqa
This amplitude    satisfies  neither the Ward identity of   the symmetric nor the antisymmetric  NSNS gauge transformations. This is again an indication of  the presence of a new T-dual multiplet  which  will be found  in  section 3.3. The combination of   the terms in the fourth, the seventh and the last lines of \reef{n=2polarization}, and   the amplitude \reef{A''1} satisfies the RR gauge transformations. One can check that all other terms are proportional to the RR momenta. They would require the   amplitude of the RR $(p+1)$-form with two transverse indices in which we are not interested in this paper, to become invariant under the RR gauge transformations.

The above amplitude  is not covariant under  T-duality along the world volume when the Killing index $y$ is carried by the NSNS polarizations. To make the above amplitude symmetric under the T-duality, one has to add the following amplitude:
\beqa
{\cal A}_4(C^{(p+3)})&\!\!\!\!\!\sim\!\!\!\!\!&  (\veps_1^{(p+3)})_{ijkl}{}^{a_2\cdots a_p}\eps_{a_0\cdots a_p}p_2^{a_1}\bigg[
2p_2^i p_3^{a_0}\bigg( (\veps_2^A)^{jk}(p_1\inn N\inn\veps_3^S)^{l} \bigg)\cI_1\nonumber\\
&&-p_2^i p_3^{a_0}\bigg( (\veps_2^A)^{jk}(p_2\inn N\inn \veps_3^S)^{l} \bigg)\cI_2 -p_3^i p_3^{a_0}\bigg( (\veps_2^A)^{jk}(p_2\inn V\inn\veps_3^S)^{l} \bigg)\cI_2\nonumber\\
&&+ p_3^i p_2\inn V\inn p_3\bigg( (\veps_2^A)^{jk}(\veps_3^S)^{a_0l} +(2\leftrightarrow 3)\bigg)\cI_2\labell{n=4polarization}\\
&&+p_2^i p_3^{a_0}\bigg( (\veps_2^A)^{jk}(p_2\inn V\inn \veps_3^S)^{l} \bigg)\cI_3 +p_3^i p_3^{a_0}\bigg( (\veps_2^A)^{jk}(p_2\inn N\inn\veps_3^S)^{l} \bigg)\cI_3\nonumber\\
&&- p_3^i p_2\inn N\inn p_3\bigg( (\veps_2^A)^{jk}(\veps_3^S)^{a_0 l} +(2\leftrightarrow 3)\bigg)\cI_3-4p_2^i p_3^{a_0}\bigg( (\veps_2^A)^{jk}(p_3\inn V\inn \veps_3^S)^{l} \bigg)\cI_4\nonumber\\
&& +2p_2^i p_3\inn V\inn p_3\bigg(  (\veps_2^A)^{jk}(\veps_3^S)^{a_0l} +(2\leftrightarrow 3)\bigg)\cI_4\bigg]+(2\leftrightarrow 3)\nonumber
\eeqa
All terms above are covariant   except  those which have $(\veps^S)^{a_0l}$. They become covariant under the T-duality when one includes the following S-matrix element for the RR $(p+5)$-form:
\beqa
{\cal A}_5(C^{(p+5)})&\!\!\!\!\!\sim\!\!\!\!\!& \frac{1}{4} (\veps_1^{(p+5)})_{ijklm}{}^{a_1\cdots a_p}\eps_{a_0\cdots a_p}p_2^{a_0}\bigg[
 - p_3^i p_2\inn V\inn p_3\bigg( (\veps_2^A)^{jk}(\veps_3^A)^{lm} +(2\leftrightarrow 3)\bigg)\cI_2\nonumber\\
&&+ p_3^i p_2\inn N\inn p_3\bigg( (\veps_2^A)^{jk}(\veps_3^A)^{ lm}+(2\leftrightarrow 3)\bigg)\cI_3\nonumber\\
&& -2p_2^i p_3\inn V\inn p_3\bigg(  (\veps_2^A)^{jk}(\veps_3^A)^{lm} +(2\leftrightarrow 3)\bigg)\cI_4\bigg]+(2\leftrightarrow 3)\labell{n=5polarization}
\eeqa
Since the NSNS polarization tensors carry no world volume indices, there is  no more S-matrix element to be added to the above amplitudes. 

Hence, the T-duality produces the following sequence: 
\beqa
 \cA_1\rightarrow \cA_2\rightarrow \cA_3\rightarrow \cA_4\rightarrow \cA_5\labell{TA}
 \eeqa
where $\cA_1,\, \cA_2,\, \cA_3,\, \cA_4$ and $\cA_5$ are given in 
 \reef{n=1polarization}, \reef{n=2polarization}, \reef{n=3polarization}, \reef{n=4polarization} and \reef{n=5polarization}, respectively. They  form the T-dual multiplet $\cA$ which is covariant under the linear T-duality transformations \reef{linear}, \ie it satisfies the T-dual Ward identity. The multiplet does not satisfy the Ward identity associated with the   the NSNS gauge transformations. In the next section we will find a new T-dual multiplet whose combination with the above multiplet is invariant under the NSNS gauge transformations.

\subsection{Gauge symmetry}

It has been shown in \cite{Garousi:2010bm} that the amplitude \reef{n=1polarization} satisfies the Ward identity associated with the   NSNS gauge transformations. The amplitude \reef{n=2polarization}, however, does not satisfy the Ward identity of the symmetric NSNS gauge transformation. So there must be another T-dual multiplet as well as the multiplet we have found in the previous section.    Taking the assumption in the  T-duality that the fields must be independent of the Killing coordinate, one realizes that the amplitudes \reef{n=2polarization}, \reef{n=3polarization}, \reef{n=4polarization} and \reef{n=5polarization} are off by some terms which are proportional to $p_2^ip_3^j$. So the new T-dual multiplet must be proportional to $p_2^ip_3^j$. 

To find the first component of this multiplet,  we impose the consistency of the amplitude \reef{n=2polarization} with the symmetric NSNS gauge transformation, \ie under replacing the symmetric NSNS polarization with
\beqa
(\veps^S)^{\mu\nu}\rightarrow p^{\mu}\z^{\nu}+p^{\nu}\z^{\mu},\labell{replace}
\eeqa
the gauge invariant amplitude must be zero. Using this condition, one finds that
  the following amplitude must be added to the amplitude \reef{n=2polarization}:
\beqa
&&{\cal A'}_{2}(C^{(p-1)}) \sim -(\veps_1^{(p-1)})_{ij}{}^{a_4\cdots a_p}\eps_{a_0\cdots a_p}p_2^ip_3^jp_2^{a_3}\bigg[
 (\veps_2^A)^{a_1a_2}\bigg(2(p_1\inn N\inn\veps_3^S)^{a_0}\cI_1
-(p_2\inn N\inn\veps_3^S)^{a_0}\cI_2\nonumber\\
&&\qquad\quad+(p_2\inn V\inn\veps_3^S)^{a_0}\cI_3-\Tr(\veps_3^S\inn D)p_3^{a_0}\cI_4\bigg)+2 p_3^{a_0} \bigg((\veps_2^S\inn V\inn\veps_3^A)^{a_2a_1}\cI_2-(\veps_2^S\inn N\inn\veps_3^A)^{a_2a_1}\cI_3\bigg)\nonumber\\
&&\qquad\quad- (\veps_3^A)^{a_0a_1}\, (p_3\inn V\inn \veps_2^S)^{a_2}  \cI_2+ (\veps_3^A)^{a_0a_1}\, (p_3\inn N\inn \veps_2^S)^{a_2}  \cI_3
%&&-2(\veps_2)^{ia_1}(\veps_3)^{a_0a_3}\bigg((p_2)^{a_2}(p_2)^j(p_3\inn V\inn p_3)\cI_4+(p_3)^{a_2}(p_3)^j(p_2\inn V\inn p_2)\cI_7\nonumber\\
%&&+(p_2\inn V\inn p_3)[(p_2)^{a_2}(p_3)^{j}\cI_2-(p_2)^j(p_3)^{a_2}\cI_3]+(p_2\inn N\inn p_3)[(p_2)^{a_2}(p_3)^{j}\cI_3-(p_2)^j(p_3)^{a_2}\cI_2]\bigg)\nonumber\\
\bigg]+(2\leftrightarrow 3)\labell{n=2polarizationadd}
\eeqa
The amplitude \reef{n=2polarization} is invariant under the   antisymmetric NSNS gauge transformations and is proportional to the RR momenta which is a necessary condition for  satisfying the Ward identity associated with the RR gauge transformation.   One can  verify that the above amplitude is also proportional to the RR momenta and satisfies the Ward identity associated with the   antisymmetric NSNS gauge transformations. 

We can do the same steps as above to find the gauge completion of the other components of the T-dual multiplet found in the previous section, or we can do the same steps as in previous section to find the T-dual completion of the above amplitude. We choose the latter method to find the T-dual multiplet corresponding to the  amplitude \reef{n=2polarizationadd}. So we consider there is isometric along the world volume direction $y$.  
The amplitude \reef{n=2polarization} is covariant  under the linear T-duality transformation when the $y$-index is carried by the RR polarization tensor, so we expect such symmetry for the above amplitude\footnote{ Apart from the last two terms in the second line of \reef{n=2polarizationadd}, it is easy to see this symmetry because the NSNS polarization tensors contract either with the volume form or with the momentum. However, the two polarizations contract with each other in the last two terms in the second line of \reef{n=2polarizationadd}. To verify that even these terms are invariant under T-duality when the $y$-index is carried by the RR potential, consider the combination of one of them and the corresponding term in the $(2\leftrightarrow 3)$ part, \ie  
\beqa
[(\veps_2^S\inn V\inn\veps_3^A)^{a_2a_1}-(\veps_3^S\inn N\inn\veps_2^A)^{a_2a_1}]\cI_2
\eeqa
Using the strategy outlined in section 2, one can easily verify that the above combination is invariant under the linear T-duality transformations \reef{linear} when the world volume indices $a_1$ and $a_2$ are not the $y$-index. Similarly for the last term in the second line of \reef{n=2polarizationadd}.}.

Even though the amplitude \reef{n=2polarizationadd} is  covariant under the T-duality when the RR potential carries the Killing index $y$, it is not covariant  when the NSNS polarizations  carry this index. Using the same steps as in the previous section, one finds that the following amplitude must be added to \reef{n=2polarizationadd}:
\beqa
&&{\cal A'}_{3}(C^{(p+1)}) \sim (\veps_1^{(p+1)})_{ijk}{}^{a_3\cdots a_p}\eps_{a_0\cdots a_p}p_2^ip_3^jp_2^{a_2}\bigg[
 2(\veps_2^S)^{a_1 k}\bigg(2(p_1\inn N\inn\veps_3^S)^{a_0}\cI_1
-(p_2\inn N\inn\veps_3^S)^{a_0}\cI_2\nonumber\\
&&\qquad\quad+(p_2\inn V\inn\veps_3^S)^{a_0}\cI_3-\Tr(\veps_3^S\inn D)p_3^{a_0}\cI_4\bigg)+(\veps_2^A)^{a_0a_1 }\bigg(2(p_1\inn N\inn\veps_3^A)^{k}\cI_1
-(p_2\inn N\inn\veps_3^A)^{k}\cI_2\nonumber\\
&&\qquad\quad+(p_2\inn V\inn\veps_3^A)^{k}\cI_3 \bigg)-2 p_3^{a_0} \bigg([(\veps_2^A\inn V\inn\veps_3^A)^{ka_1}+(\veps_2^S\inn V\inn\veps_3^S)^{a_1 k}]\cI_2-[(\veps_2^A\inn N\inn\veps_3^A)^{ka_1}\nonumber\\
&&\qquad\quad+(\veps_2^S\inn N\inn\veps_3^S)^{a_1 k}]\cI_3\bigg) - [(\veps_3^A)^{a_0a_1}\, (p_3\inn V\inn \veps_2^A)^{k}-  2(\veps_3^S)^{a_0 k}\, (p_3\inn V\inn \veps_2^S)^{a_1}]\cI_2\nonumber\\
&&\qquad\quad+ [(\veps_3^A)^{a_0a_1}\, (p_3\inn N\inn \veps_2^A)^{k} -2(\veps_3^S)^{a_0 k}\, (p_3\inn N\inn \veps_2^S)^{a_1}] \cI_3
%&&-2(\veps_2)^{ia_1}(\veps_3)^{a_0a_3}\bigg((p_2)^{a_2}(p_2)^j(p_3\inn V\inn p_3)\cI_4+(p_3)^{a_2}(p_3)^j(p_2\inn V\inn p_2)\cI_7\nonumber\\
%&&+(p_2\inn V\inn p_3)[(p_2)^{a_2}(p_3)^{j}\cI_2-(p_2)^j(p_3)^{a_2}\cI_3]+(p_2\inn N\inn p_3)[(p_2)^{a_2}(p_3)^{j}\cI_3-(p_2)^j(p_3)^{a_2}\cI_2]\bigg)\nonumber\\
\bigg]+(2\leftrightarrow 3)\labell{add1}
\eeqa
The  amplitude is proportional to  the RR momenta which is inherited from the amplitude \reef{n=2polarizationadd}. It does not satisfy the Ward identities corresponding to the NSNS gauge transformations. However, the combination of the above amplitude and the amplitude \reef{n=3polarization} satisfies these identities.

The amplitude \reef{add1} does not satisfy the T-dual Ward identity   unless one adds to it the following amplitude:
\beqa
&&{\cal A'}_{4}(C^{(p+3)}) \sim -(\veps_1^{(p+3)})_{ijkl}{}^{a_2\cdots a_p}\eps_{a_0\cdots a_p}p_2^ip_3^jp_2^{a_1}\bigg[
 (\veps_2^A)^{ kl}\bigg(2(p_1\inn N\inn\veps_3^S)^{a_0}\cI_1
-(p_2\inn N\inn\veps_3^S)^{a_0}\cI_2\nonumber\\
&&\qquad\quad+(p_2\inn V\inn\veps_3^S)^{a_0}\cI_3-\Tr(\veps_3^S\inn D)p_3^{a_0}\cI_4\bigg)-2(\veps_2^S)^{a_0 k}\bigg(2(p_1\inn N\inn\veps_3^A)^{l}\cI_1
-(p_2\inn N\inn\veps_3^A)^{l}\cI_2\nonumber\\
&&\qquad\quad+(p_2\inn V\inn\veps_3^A)^{l}\cI_3 \bigg) -2 p_3^{a_0} \bigg((\veps_2^A\inn V\inn\veps_3^S)^{kl} \cI_2-(\veps_2^A\inn N\inn\veps_3^S)^{kl}\ \cI_3\bigg)\nonumber\\
&&\qquad\quad - \bigg(2(\veps_3^S)^{a_0 l}\, (p_3\inn V\inn \veps_2^A)^{k}+  (\veps_3^A)^{ kl}\, (p_3\inn V\inn \veps_2^S)^{a_0}\bigg)\cI_2\nonumber\\
&&\qquad\quad+ \bigg(2(\veps_3^S)^{a_0 l}\, (p_3\inn N\inn \veps_2^A)^{k} +(\veps_3^A)^{ kl}\, (p_3\inn N\inn \veps_2^S)^{a_0}\bigg) \cI_3
%&&-2(\veps_2)^{ia_1}(\veps_3)^{a_0a_3}\bigg((p_2)^{a_2}(p_2)^j(p_3\inn V\inn p_3)\cI_4+(p_3)^{a_2}(p_3)^j(p_2\inn V\inn p_2)\cI_7\nonumber\\
%&&+(p_2\inn V\inn p_3)[(p_2)^{a_2}(p_3)^{j}\cI_2-(p_2)^j(p_3)^{a_2}\cI_3]+(p_2\inn N\inn p_3)[(p_2)^{a_2}(p_3)^{j}\cI_3-(p_2)^j(p_3)^{a_2}\cI_2]\bigg)\nonumber\\
\bigg]+(2\leftrightarrow 3)\labell{add2}
\eeqa
The terms in the big bracket which have $p_3^{a_0}$ are covariant under the T-duality. All  other terms which are not covariant, dictate that the whole amplitude must include the following amplitude for the RR $(p+5)$-form:
\beqa
&&{\cal A'}_{5}(C^{(p+5)}) \sim (\veps_1^{(p+5)})_{ijklm}{}^{a_1\cdots a_p}\eps_{a_0\cdots a_p}p_2^ip_3^jp_2^{a_0}\bigg[
 (\veps_2^A)^{ kl}\bigg(2(p_1\inn N\inn\veps_3^A)^{m}\cI_1
-(p_2\inn N\inn\veps_3^A)^{m}\cI_2\nonumber\\
&&\qquad+(p_2\inn V\inn\veps_3^A)^{m}\cI_3 \bigg)   - (\veps_3^A)^{ lm}\, (p_3\inn V\inn \veps_2^A)^{k} \cI_2 +  (\veps_3^A)^{ lm}\, (p_3\inn N\inn \veps_2^A)^{k}   \cI_3
%&&-2(\veps_2)^{ia_1}(\veps_3)^{a_0a_3}\bigg((p_2)^{a_2}(p_2)^j(p_3\inn V\inn p_3)\cI_4+(p_3)^{a_2}(p_3)^j(p_2\inn V\inn p_2)\cI_7\nonumber\\
%&&+(p_2\inn V\inn p_3)[(p_2)^{a_2}(p_3)^{j}\cI_2-(p_2)^j(p_3)^{a_2}\cI_3]+(p_2\inn N\inn p_3)[(p_2)^{a_2}(p_3)^{j}\cI_3-(p_2)^j(p_3)^{a_2}\cI_2]\bigg)\nonumber\\
\bigg]+(2\leftrightarrow 3)\labell{add3}
\eeqa
There are no more world volume indices for the NSNS polarization tensors, so there is no more amplitude to be added for the consistency with the linear T-duality. 

Therefore,  the sequence of the amplitudes  
\beqa
   \cA'_2\rightarrow \cA'_3\rightarrow \cA'_4\rightarrow \cA'_5\labell{TA'}
 \eeqa
where $  \cA'_2,\, \cA'_3,\, \cA'_4$ and $\cA'_5$ are given in   \reef{n=2polarizationadd}, \reef{add1}, \reef{add2} and \reef{add3}, respectively,   forms the T-dual multiplet $\cA'$. This multiplet is related to the multiplet in the previous section through the gauge transformations, \ie
 \beqa
\matrix{\cA_1& \!\!\!\!\!\rightarrow \!\!\!\!\!& \cA_2& \!\!\!\!\!\rightarrow \!\!\!\!\!&\cA_3&  \!\!\!\!\!\rightarrow \!\!\!\!\! & \cA_4& \!\!\!\!\!\rightarrow \!\!\!\!\!&\cA_5\cr 
&&\downarrow & &\downarrow&& \downarrow&& \downarrow\cr
&&  \cA'_2& \!\!\!\!\!\rightarrow \!\!\!\!\!&\cA'_3& \!\!\!\!\!\rightarrow \!\!\!\!\!&\cA'_4& \!\!\!\!\!\rightarrow \!\!\!\!\!&\cA'_5 }
 \labell{zero2nz} 
 \eeqa 
 where the pairs $\cA_i(\cA'_i)$ and $\cA_{i+1}(\cA'_{i+1})$ related through the T-duality are connected by horizontal arrows, and pairs  $\cA_i$ and $\cA'_i$ related through the gauge invariance are connected by vertical arrows.
Neither the T-dual multiplet ${\cal A}$   nor the   multiplet ${\cal A'}$, satisfy the Ward identity corresponding to the NSNS gauge transformations, however, the combination of these two multiplets, \ie
\beqa
\textbf{A}&\equiv&{\cal A} +{\cal A}'\labell{AA}
\eeqa
satisfies the Ward identity. So   they should be the correct S-matrix elements of one RR and two NSNS vertex operators. While there is no restriction on the NSNS polarization tensors, the RR potential carries  specific indices. That is, the $(p-3)$-form, $(p-1)$-form, $(p+1)$-form,  $(p+3)$-form and $(p+5)$-form carry one, two, three, four and five transverse indices, respectively. 
 
 %\subsection{Explicit calculations}

To compare our result with explicit calculation in string theory, recall that  the amplitude \reef{n=1polarization} corresponds to the first term of the sequence \reef{seq} when the RR potential has one transverse index \cite{Garousi:2010bm}. The trace \reef{trace} in this case is nonzero for $m=6$. All other components of the T-dual multiplets \reef{AA} should   correspond to the other terms of the   sequence \reef{seq} for $m=6$, \ie
 \beqa
 \textbf{A}& \stackrel{?}{\sim}&T_p\delta^{p+1}(p_1^a+p_2^a+p_3^a)\sum_{j=0}^{4} \sum T(p-3+2j,p,6)\cI \,\labell{rel.2}
 \eeqa
 where  $\cI$ stands for the functions of the Mandelstam variables which appear in the amplitude. The first term in the first summation    has one transverse indices, the second term has two transverse indices and so on.   For $m=6$ there are  many  $A_{[\alpha_1\cdots \alpha_6]}$ in the trace \reef{trace}. The second summation  in \reef{rel.2} is over these terms. Extending the calculations done in \cite{Garousi:2010bm} for the first term of the sequence \reef{seq}, to all other terms, we have computed explicitly the right hand side of the above equation and found exact agreement with the T-dual multiplets in the left hand side of \reef{rel.2}.

\subsection{Amplitudes in terms of field strength $H$}

The amplitudes that we have found in the previous section satisfy the Ward identity associated with the NSNS gauge transformations. So one may use these symmetries to rewrite the amplitudes in terms of the field strength of the   NSNS states. The amplitude \reef{n=1polarization} has been written in terms of $H$ in \cite{Garousi:2010bm}, \ie  
\beqa
\textbf{A}_1(C^{(p-3)})&\sim& (\veps_1^{(p-3)})_i{}^{a_5\cdots a_p}\eps_{a_0\cdots a_p}H_2^{a_1a_4a_2}\bigg( -2(p_1\inn N\inn H_3)^{a_3a_0}p_2^{i}\cI_1+4(p_3\inn V\inn H_3)^{a_3a_0}p_2^{i}\cI_4
\nonumber\\
&&\qquad\qquad\qquad-(p_2\inn W\inn H_3)^{a_3a_0}p_2^{i}
+(p_2\inn U\inn H_3)^{a_3a_0}p_3^{i} 
\bigg) +(2\leftrightarrow 3)\labell{totalf}
\eeqa
where $H_i$ for $i=2,3$, is the field strength of the $B_i$ polarization tensor, \eg
\beqa
(H_2)^{\mu\nu\alpha}&=&i[\veps_2^{\mu\nu}p_2^{\alpha}+\veps_2^{\alpha\mu}p_2^{\nu}+\veps_2^{\nu\alpha}p_2^{\mu}]
%R_{\mu\nu\gamma\lambda}&=&\frac{1}{2}\bigg(h_{\mu\lambda,\nu\gamma}+h_{\nu\gamma,\mu\lambda}-h_{\mu\gamma,\nu\lambda}-h_{\nu\lambda,\mu\gamma}\bigg)\nonumber\\
%\hat{R}_{\mu\nu}&=&R_{\mu a\nu a}\nonumber
\eeqa
In converting  the result in \cite{Garousi:2010bm} to the above form,   we have  defined $U$ and $W$ as
\beqa
U\equiv V\cI_2-N\cI_3&;&W\equiv V\cI_3-N\cI_2
\eeqa
Apart from the first term in \reef{totalf}, no other term has  $p_1$ so the amplitude  can not be written in terms of the RR field strength unless one rewrite it in terms of the B-field polarizations instead of the field strengths $H_2, H_3$. However, one expects the contact terms of the amplitude \reef{totalf} to be combined with  some other contact terms   to produce couplings in terms of field strength of the external states. We will study this issue in the next section.

The next component of the T-dual multiplet $\textbf{A}$ has the two contributions \reef{n=2polarization} and \reef{n=2polarizationadd}. The amplitude is nonzero when one of the NSNS states is symmetric and the other one is antisymmetric. Assuming the polarization $\veps_2$ is symmetric and the polarization $\veps_3$ is antisymmetric, one can rewrite the amplitude in terms of $H_3$ as
 \beqa
\textbf{A}_2(C^{(p-1)})&\sim &2(\veps_1^{(p-1)})_{ij}{}^{a_4\cdots a_p}\eps_{a_0\cdots a_p}\bigg[-H_3^{a_0a_1a_2}p_3^{j}\bigg(2p_2^{[i}(p_1\inn N\inn \veps_2^S)^{a_3]} \cI_1+p_2^{[i}(p_3\inn U\inn \veps_2^S)^{a_3]}  \nonumber\\
&& +[4p_2^{a_3}(p_2\inn V\inn\veps_2^S)^i-p_2^{a_3}p_2^i \Tr(\veps_2^S\inn D)-2p_2\inn V\inn p_2(\veps_2^S)^{a_3i}]\cI_7\bigg)\nonumber\\
&&-3p_2^ip_2^{a_3}(\veps_2^S)^{a_2j}\bigg(2(p_1\inn N\inn H_3)^{a_0a_1}\cI_1 +(p_2\inn W\inn H_3)^{a_0a_1}  -4(p_3\inn V\inn H_3)^{a_0a_1}\cI_4\bigg)\nonumber\\
&&+3p_3^{j}p_2^{a_3}\bigg(-(\veps_2^S)^{a_2i} (p_2\inn U\inn H_3)^{a_0a_1} +p_2^i(\veps_2^S \inn U\inn H_3)^{a_0a_1a_2}  \bigg)\nonumber\\
&&+ H_3^{a_0a_1a_2}p_2^j\bigg((\veps_2^S)^{ia_3}p_2\inn W\inn p_3\ -p_2^{a_3}(p_3\inn W\inn\veps_2^S)^i \bigg) \bigg]\labell{totalff}
\eeqa
Note that we have used the identity \reef{identity} to write the above amplitude in terms of field strength $H_3$. The amplitude does not satisfy the Ward identity corresponding to the graviton unless one rewrite the field strength $H_3$ in terms of $\veps_3$. So one can not write the amplitude in terms of  field strengths  $H_3$ and the curvature $R_2$. However, a part of it  can be written as $R_2H_3$. In fact one does not expect that even all the contact terms to be rewritten as $R_2H_3$ because the metric   in the effective action appears in the curvature tensor, in contracting the indices and      in the definition of the covariant derivatives.  

The $C^{(p+1)}$ component of the multiplet $\textbf{A}$ is nonzero when both NSNS polarizations are either antisymmetric or symmetric. In the former case, the amplitude in terms of $H$ is
 \beqa
\textbf{A}^A_3(C^{(p+1)})&\sim & (\veps_1^{(p+1)})_{ijk}{}^{a_3\cdots a_p}\eps_{a_0\cdots a_p}\bigg[p_2^kH_2^{a_2a_1a_0}\bigg(2(p_1\inn N\inn H_3)^{ij}\cI_1+(p_2\inn W\inn H_3)^{ij} \bigg)\nonumber\\
&&+p_2^{a_2}H_2^{kji}\bigg(2(p_1\inn N\inn H_3)^{a_1a_0}\cI_1+(p_2\inn W\inn H_3)^{a_1a_0} -4(p_3\inn V\inn H_3)^{a_1a_0}\cI_4\bigg)\nonumber\\
&& + 3p_3^kp_2^{a_2}(H_2\inn U\inn H_3)^{ija_1a_0} +\frac{1}{3}H_2^{kji}H_3^{a_2a_1a_0} (p_2\inn W\inn p_3) \bigg)\bigg]+(2\leftrightarrow 3)\labell{totalfffA}
\eeqa
In the latter case the amplitude in terms of the symmetric polarization tensors is 
 \beqa
\textbf{A}^S_3(C^{(p+1)})&\!\!\sim\!\!\!& 12(\veps_1^{(p+1)})_{ijk}{}^{a_3\cdots a_p}\eps_{a_0\cdots a_p}\bigg[p_2^{a_0}p_2^j(\veps_2^S)^{a_1k}\bigg(2(p_1\inn N\inn \veps_3^S)^{[a_2}p_3^{i]}\cI_1 +(p_2\inn W\inn \veps_3^S)^{[a_2}p_3^{i]}\nonumber\\
&& +[4(p_3\inn V\inn \veps_3^S)^ip_3^{a_2}-p_3^{a_2}p_3^i\Tr(\veps_3^S\inn D)-2p_3\inn V\inn p_3(\veps_3^S)^{a_2i}]\cI_4\bigg)\nonumber\\
&&+p_3^{a_1}p_2^i\bigg((\veps_2^S)^{ja_0}(\veps_3^S)^{ka_2}p_2\inn W\inn p_3 -p_3^k(\veps_2^S)^{ja_0}(p_2\inn W\inn \veps_3^S)^{a_2}   \nonumber\\
&&-p_2^{a_0}(\veps_3^S)^{ka_2}(p_3\inn W\inn \veps_2^S)^j+p_3^kp_2^{a_0}(\veps_2^S\inn W\inn \veps_3^S)^{ja_2} \bigg)\bigg] +(2\leftrightarrow 3)\labell{totalfffS}
\eeqa
In this case also one can not rewrite the amplitude   in terms of curvature tensors $R_2R_3$, even though a part of it  can be.

 The next component of the T-dual multiplet $\textbf{A}$ is  nonzero when one of the NSNS states is symmetric and the other one is antisymmetric. Assuming the polarization $\veps_2$ is symmetric and the polarization $\veps_3$ is antisymmetric, one can rewrite the amplitude in terms of $H_3$ as
 \beqa
\textbf{A}_4(C^{(p+3)})&\sim &2(\veps_1^{(p+3)})_{ijkl}{}^{a_2\cdots a_p}\eps_{a_0\cdots a_p}\bigg[H_3^{klj}p_3^{a_0}\bigg(2p_2^{[i}(p_1\inn N\inn \veps_2^S)^{a_1]} \cI_1+p_2^{[i}(p_3\inn U\inn \veps_2^S)^{a_1]} \nonumber\\
&&+[4p_2^{a_1}(p_2\inn V\inn\veps_2^S)^i-p_2^{a_1}p_2^i\Tr(\veps_2^S\inn D)-2p_2\inn V\inn p_2(\veps_2^S)^{a_1i}]\cI_7\bigg)\nonumber\\
&&+3p_2^{a_0}p_2^i(\veps_2^S)^{a_1j}\bigg(2(p_1\inn N\inn H_3)^{kl}\cI_1 +(p_2\inn W\inn H_3)^{kl} \bigg)\nonumber\\
&&+3p_3^{a_0}p_2^i\bigg(p_2^{a_1}(\veps_2^S\inn W\inn H_3)^{jlk} -(\veps_2^S)^{ja_1}(p_2\inn W\inn H_3)^{lk} \bigg)\nonumber\\
&&+ H_3^{klj}p_2^{a_0}\bigg(-(\veps_2^S)^{a_1i}p_2\inn U\inn p_3 +p_2^i(p_3\inn U\inn \veps_2^S)^{a_1} \bigg) \bigg] \labell{totalffff}
\eeqa
Note that  in using the identity \reef{identity} to write the amplitudes \reef{n=4polarization} and \reef{add2} in terms of $H_3$, the $\cI_4$ terms disappear. This is  unlike the amplitude \reef{totalff}. As we shall see in section 4,  this causes to have a low energy contact term at order $\alpha'^2$ for the amplitude \reef{totalff} but no such contact term for the amplitude \reef{totalffff}.

The last component is nonzero when both NSNS polarization tensors are antisymmetric. The amplitude in terms of $H$ is the following:
 \beqa
\textbf{A}_5(C^{(p+5)})&\sim& (\veps_1^{(p+5)})_{ijklm}{}^{a_1\cdots a_p}\eps_{a_0\cdots a_p}H_2^{mkl}\bigg( 2(p_1\inn N\inn H_3)^{ij}p_2^{a_0}\cI_1\nonumber\\&&+(p_2\inn N\inn H_3)^{ij}p_1^{a_0}\cI_2 + (p_2\inn V\inn H_3)^{ij}p_1^{a_0}\cI_3
\bigg)+(2\leftrightarrow 3)\labell{fffff} 
\eeqa
Note that  there is no $\cI_4$ term in this amplitude either, whereas there is such term in the first component of the multiplet $\textbf{A}$, \ie the amplitude \reef{totalf}. As we shall see in   section 4, this leads to a  contact term at order $\alpha'^2$ in the amplitude \reef{totalf} and no such contact term in the above amplitude.

\subsection{Two closed and one open string amplitudes}

To study the low energy effective action of the amplitudes we have found in the previous sections, we need to have the T-dual multiplet corresponding to the  S-matrix elements of one RR and one NSNS states which has been found in \cite{ Garousi:1996ad,Garousi:2010ki}, and the T-dual multiplet corresponding to the S-matrix elements of one RR, one NSNS and one open string NS states. The first component of this latter multiplet   has been calculated in \cite{Becker:2011ar}. For the case that the RR potential has one transverse index, the amplitude is 
\beqa
A_1(C^{(p-3)})&\sim&T_p(\veps_1^{(p-3)})_i{}^{a_5\cdots a_p}\eps_{a_0\cdots a_p}p_3^ip_3^{a_4}(\veps_3^A)^{a_3a_2}f_2^{a_0a_1}{\cal Q}\labell{A1}
\eeqa
where ${\cal Q}$ is the following  function of the Mandelstam variables \cite{Becker:2011ar}:
\beqa
{\cal Q}&=&\frac{1}{p_1\inn p_3}\pmatrix{p_1\inn p_3+p_3\inn D\inn p_3 \cr 
p_1\inn p_3 }^{-1}
\eeqa
and $f_2^{a_0a_1}=i[k_2^{a_0}\z_2^{a_1}-k_2^{a_1}\z_2^{a_0}]$. This amplitude is covariant under the linear T-duality when the isometry index  $y$ is carried by the RR potential. However, when the $y$-index is carried by the NS or the  NSNS polarizations it is not invariant. Using the same steps as we have done in section 3.1, one finds that  the following amplitudes have to be added to the above amplitude to have complete symmetry under the linear T-duality transformations:
\beqa
A_2(C^{(p-1)})&\sim&2T_p(\veps_1^{(p-1)})_{ij}{}^{a_4\cdots a_p}\eps_{a_0\cdots a_p}p_3^ip_3^{a_3}[(\veps_3^S)^{a_2j}f_2^{a_0a_1}+i(\veps_3^A)^{a_1a_2}k_2^{a_0}\Phi^j_2]{\cal Q}\nonumber\\
A_3(C^{(p+1)})&\sim&T_p(\veps_1^{(p+1)})_{ijk}{}^{a_3\cdots a_p}\eps_{a_0\cdots a_p}p_3^ip_3^{a_2}[-(\veps_3^A)^{jk}f_2^{a_0a_1}+4i(\veps_3^S)^{a_1j}k_2^{a_0}\Phi^k_2]{\cal Q}\nonumber\\
A_4(C^{(p+3)})&\sim&2iT_p(\veps_1^{(p+3)})_{ijkl}{}^{a_2\cdots a_p}\eps_{a_0\cdots a_p}p_3^ip_3^{a_1}k_2^{a_0}(\veps_3^A)^{jk} \Phi^l_2{\cal Q}\labell{A2}
\eeqa
  The amplitudes \reef{A1} and \reef{A2} form a T-dual multiplet which satisfies all the Ward identities.

\section{Low energy effective field theory }

The dynamics of the D-branes of type II superstring theories at the lowest order in $\alpha'$ is  given by the   world-volume theory which is the  sum of   Dirac-Born-Infeld (DBI) and Chern-Simons (CS) actions  \cite{Leigh:1989jq,Bachas:1995kx,Polchinski:1995mt,Douglas:1995bn}. The bosonic part of this action is  
\beqa
S&=&-T_p\int d^{p+1}x\,e^{-\phi}\sqrt{-\det\left(G_{ab}+B_{ab}\right)}+T_{p}\int_{M^{p+1}}e^{B}C+\cdots\labell{DBI}
\eeqa
here dots refer to the   terms at higher order of $\alpha'$. In the Chern-Simos part, $M^{p+1}$ represents the world volume of the D$_p$-brane, $C$ is the sum over all  RR potential forms, \ie $C=\sum_{n=0}^8C^{(n)}$,  and the multiplication rule is  the wedge product. It is understood here that after expanding the exponential, in each term that particular $C^{(n)}$ is chosen  such that the product of  forms adds up to a $(p+1)$-form. The field strength of the four form  is self-dual object  and the electric components of the redundant fields $C^{(5)},\cdots, C^{(8)}$ are related to the magnetic components of the RR fields $C^{(0)},\cdots, C^{(3)}$ as $dC^{(8-n)}=*(dC^{(n)})$ for $n=0,1,2,3$. The closed string fields $G_{ab}, B_{ab}$ and $C$ are  the pull-back of the bulk fields   onto the world-volume of D-brane, \eg
\beqa
G_{ab}&=&G_{\mu\nu}\frac{\partial x^{\mu}}{\prt\sigma^a}\frac{\prt x^{\nu}}{\prt\sigma^b}
\eeqa
where $x^i=2\pi\alpha' \Phi^i$ in the static gauge.  The transverse scalar fields appear in the D-brane action also in the Taylor expansion of the closed string fields \cite{Garousi:1998fg}. The abelian gauge field can be added to the action as $B\rightarrow B+2\pi\alpha'f$. This causes the action to be invariant under the B-field gauge transformation. The Chern-Simons part is invariant under the following RR gauge transformations:
\beqa
\delta C=d\Lambda+H\wedge \Lambda \labell{RR}
\eeqa
  where $H$ is the field strength of $B$, \ie $H=dB$, and $\Lambda=\sum_{n=0}^7\Lambda^{(n)}$.

The bosonic part of the bulk action at the lowest order in $\alpha'$  for type IIA theory is given as (see \eg \cite{Bergshoeff:1995as}):
\beqa
S_{IIA}&=&\frac{1}{2\kappa ^2}\int d^{10}x\sqrt{-G}\bigg(e^{-2\phi}\bigg[R+4(\nabla\phi)^2-\frac{1}{12}H^2\bigg]-\frac{1}{2n!}(\tilde{F}^{(n)})^2\bigg)\nonumber\\
&&-\frac{1}{4\kappa ^2}\int B\wedge dC^{(3)}\wedge dC^{(3)}+\cdots\labell{IIA}
\eeqa
where  the degree of the nonlinear RR field strength $\tilde{F}^{(n)}$ is $n=2,4$. The ten-dimensional Newton's constant is given by $2\kappa^2=16\pi G_N=(2\pi)^7\alpha'^4g_s^2$. The action for the type IIB theory is
\beqa
S_{IIB}&=&\frac{1}{2\kappa ^2}\int d^{10}x\sqrt{-G}\bigg(e^{-2\phi}\bigg[R+4(\nabla\phi)^2-\frac{1}{12}H^2\bigg]-\frac{1}{2\alpha n!}(\tilde{F}^{(n)})^2\bigg)\nonumber\\
&&+\frac{1}{4\kappa ^2}\int (C^{(4)}+\frac{1}{2}B\wedge C^{(2)})\wedge dC^{(2)}\wedge H+\cdots\labell{IIB}
\eeqa
where $n=1,3,5$, the constant $\alpha $ is 1 for $n=1,3$ and is 2 for $n=5$. The dots in above actions again refer to the higher derivative terms. The nonlinear RR field strength in above actions is given as
\beqa
\tilde{F}^{(n)}=dC^{(n-1)}+H\wedge C^{(n-3)}\labell{tF}
\eeqa
The self duality constraint $\tilde{F}^{(5)}=*\tilde{F}^{(5)}$ is imposed by hand \cite{Bergshoeff:1995sq}. In writing the above supergravity actions, we have used the conventions used in \cite{Myers:1999ps} for the RR field strength which  coincide with the RR potentials appearing in the D-brane action \reef{DBI}, \ie the supergravity actions and the D-brane action  are invariant under the same RR gauge transformations \reef{RR}. 

Using the above actions one may compute the scattering amplitude of one RR and two NSNS states. They should coincide with the amplitudes that we have found in the previous section at the lowest order in $\alpha'$. The next to the leading order terms of the string theory amplitudes in which we are interested in this section, should reproduced by the $\alpha'^2$  corrections to the above actions. It is known that the first $\alpha'$ corrections to the type II supergravity actions are at order $\alpha'^3$ \cite{Green:1981ya,Grisaru:1986dk,Grisaru:1986px,Gross:1986iv}.   So the $\alpha'^2$ corrections appear only for the D-brane action. 

The $\alpha'^2$ corrections for the couplings of one RR and one NSNS states have been found in \cite{Garousi:2010ki}. That quadratic  couplings can be extended to higher order terms by making them   covariant under the coordinate transformations and invariant under the RR gauge transformations \reef{RR}. That is, the partial derivatives  in these couplings should be replaced by the covariant derivatives, the closed string tensors should be extended to the pull-back of the bulk fields   onto the world-volume of D-brane,  the linear RR gauge field strength   should be extended to the nonlinear field strength   $\tilde{F}^{(n)}$ given in \reef{tF}, and the linear curvature tensor should be extended to the nonlinear curvature tensor $R$, \ie  
\beqa
S&\!\!\!\!\!\supset\!\!\!\!&\pi^2\alpha'^2T_p\int d^{p+1}x\,\eps^{a_0\cdots a_p}\left(-\frac{1}{2!(p-1)!}[{ \tilde{F}}^{(p)}_{ia_2\cdots a_p}H_{a_0a_1a}{}^{;ai}+{ \tilde{F}}^{(p)}_{aa_2\cdots a_p,i}H_{a_0a_1}{}^{i;a}]\right.\nonumber\\
&&\left.\qquad\qquad\qquad\qquad-\frac{2}{p!}[\frac{1}{2!}{\tilde{F}}^{(p+2)}_{ia_1\cdots a_pj}R_{aa_0}{}^{ij;a}+\frac{1}{p+1}{ \tilde{F}}^{(p+2)}_{a_0\cdots a_pj;i}( R_a{}^{iaj}-\phi\,^{;ij})]\right. \nonumber\\
&&\left.\qquad\qquad\qquad\qquad+\frac{1}{3!(p+1)!}{ \tilde{F}}^{(p+4)}_{ia_0\cdots a_pjk}H^{ijk;a}{}_a\right)\labell{first111}
\eeqa
where the semicolons   are used to denote the covariant differentiation.

The $\alpha'^2$ corrections to the D-brane action for one RR, one NSNS and one open string NS states can be found from the low energy limit of the T-dual multiplet found in section 3.5. The low energy expansion of the function ${\cal Q}$ in this multiplet is \cite{Becker:2011ar}
\beqa
{\cal Q}&=&\frac{1}{p_1\inn p_3}-\frac{\pi^2}{6}p_3\inn D\inn p_3+\cdots
\eeqa
The first term is a closed string pole which is reproduced by the D-brane action \reef{DBI} and the supergravity actions. The second term produces the following $\alpha'^2$ couplings on the world volume of the  D$_p$-brane:
\beqa
&&\pi^2\alpha'^2T_p\int d^{p+1}x\,\eps^{a_0\cdots a_p}\left(\frac{1}{2!2!(p-3)!}[{ F}^{(p-2)}_{ia_4\cdots a_p}H_{a_0a_1a}{}^{,ai}( 2\pi\alpha'f_{a_2a_3}) ]\right.\nonumber\\
&&\left.\qquad\qquad+ \frac{1}{2!(p-2)!}{ F}^{(p)}_{ija_3\cdots a_p}[\cR_{aa_0}{}^{ij,a}( 2\pi\alpha'f_{a_1a_2})-H_{a_0a_1a}{}^{,ai}(2\pi\alpha'\Phi^{j}{}_{,a_2})]\right. \nonumber\\
&&\left.\qquad\qquad+\frac{1}{(p-1)!}{ F}^{(p+2)}_{ika_2\cdots a_pj}[\frac{1}{2!3!}H^{ijk,a}{}_a( 2\pi\alpha'f_{a_0a_1})-\cR_{aa_0}{}^{ij,a}(2\pi\alpha'\Phi^{k}{}_{,a_1})]\right. \nonumber\\
&&\left.\qquad\qquad+\frac{1}{3!p!}{ F}^{(p+4)}_{ila_1\cdots a_pjk}H^{ijk,a}{}_a(2\pi\alpha'\Phi^{l}{}_{,a_0})\right)\labell{first112}
\eeqa
where $\cR$ is the linear curvature tensor and $F^{(n)}=dC^{(n-1)}$. 
One can easily verify that the couplings which involve the transverse scalar fields, are exactly reproduced by the pull-back operator in \reef{first111} in the static gauge. The other couplings are new couplings which should be added to the D-brane action at order $\alpha'^2$. Extending them to be covariant under the coordinate transformations and invariant under the   RR gauge transformations, one finds the following nonlinear couplings at order $\alpha'^2$:
\beqa
S&\!\!\!\!\!\supset\!\!\!\!&\pi^2\alpha'^2T_p\int d^{p+1}x\,\eps^{a_0\cdots a_p}\left(\frac{1}{2!2!(p-3)!}[{\tilde{F}}^{(p-2)}_{ia_4\cdots a_p}H_{a_0a_1a}{}^{;ai}(B_{a_2a_3}+2\pi\alpha'f_{a_2a_3}) ]\right.\nonumber\\
&&\left.\qquad\qquad+ \frac{1}{2!(p-2)!}{\tilde{F}}^{(p)}_{ija_3\cdots a_p}[R_{aa_0}{}^{ij;a}(B_{a_1a_2}+2\pi\alpha'f_{a_1a_2}) ]\right. \nonumber\\
&&\left.\qquad\qquad+\frac{1}{(p-1)!}{\tilde{F}}^{(p+2)}_{ika_2\cdots a_pj}[\frac{1}{2!3!}H^{ijk;a}{}_a(B_{a_0a_1}+2\pi\alpha'f_{a_0a_1}) ] \right)\labell{first113}
\eeqa
where we have also extended $2\pi\alpha'f$ to gauge invariant combination $B+2\pi\alpha'f$.

 Using the above actions one can calculate the Feynman amplitudes for the scattering of one RR and two NSNS states at order $\alpha'^2$ and compare the results with the T-dual multiplet $\textbf{A}$  at order $\alpha'^2$.  The Feynman amplitude of one RR $(p-3)$-form with  one transverse index, and two arbitrary B-fields has been studied in \cite{Garousi:2010bm}. It has been shown that the open string pole results from the coupling in the first line of \reef{first113} and the coupling $Bf$ in the DBI action \reef{DBI}, and the contact terms in the first line of \reef{first113} and in the first term of \reef{first111}, produce exactly the open string pole in the amplitude \reef{totalf}. Here we perform similar computation for other components of the  multiplet $\textbf{A}$. 
 
 \subsection{$C^{(p-1)}$ case}
 
To simplify the calculation in field theory, we assume the RR potentials carry no world volume indices. So to study the component \reef{totalff} of the T-dual multiplet $\textbf{A}$ we have to set $p=3$.  The field theory amplitude is nonzero when one of the NSNS states is symmetric and the other one is antisymmetric. It is convenient to  separate the Feynman diagrams in field theory into two classes $(i)$ and $(ii)$. The class $(i)$ includes  the contact terms and the open string  pole in which the gauge boson propagates. The class $(ii)$ includes all the closed string poles and the open string pole in which the transverse scalar fields  propagate. Our reason for this classification is that, as we will we shortly, the low energy expansion of the functions $\cI$'s in the string amplitude have only massless open and closed string poles. The amplitudes in  class $(i)$ combines to produce the appropriate massless open string poles in the string theory side.  

The Feynman diagrams in the class $(i)$ are given in the figure 1. \\

 \begin{center} \begin{picture}(150,100)(0,0)
\SetColor{Black}
\Line(10,40)(10,90)
\Line(10,90)(30,110)
\Line(30,110)(30,60)
\Line(30,60)(10,40)
\Text(0,108)[]{$D_3$-brane}
\Vertex(20,90){1.5}\Vertex(20,60){1.5}
\Photon(20,60)(20,90){3}{8} \Text(12,75)[]{$A$}
\Gluon(20,90)(65,90){3}{4} \Text(42.5,102.5)[]{$B_3$}
\Gluon(20,60)(65,60){2}{10}\Text(42.5,70)[]{$h_2$}
\Photon(20,60)(50,35){3}{4} \Text(60,45)[]{$C_1^{(ij)}$}
%\Text(20,20)[]{(1.a)}
\end{picture}
%\\{\sl Figure 1 :{\rm Feynman diagram for massless open string pole.}}
%\end{center}
%\begin{center}  
\begin{picture}(50,0)(0,0)
\SetColor{Black}
\Line(10,40)(10,90)
\Line(10,90)(30,110)
\Line(30,110)(30,60)
\Line(30,60)(10,40)
\Text(0,108)[]{$D_3$-brane}
%\Text(25,20)[]{(1.b)}
\Vertex(20,75){1.5}%\Vertex(20,60){1.5}
%\Photon(20,60)(20,90){1}{8} \Text(14,75)[]{$\Phi$}
\Gluon(20,75)(65,74){2}{10} \Text(63,83)[]{$h_2$}
\Gluon(20,75)(65,97.5){3}{5}\Text(60,105)[]{$B_3$}
\Photon(20,75)(55,50){3}{5} \Text(66.5,56)[]{$C_1^{(ij)}$}
\end{picture}
\end{center}
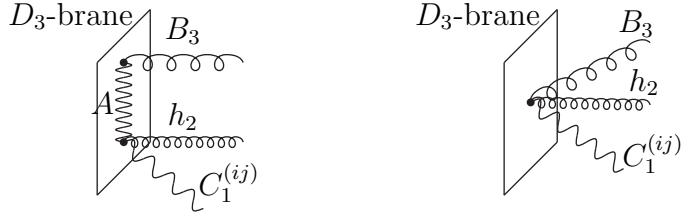
{\sl Figure 1 :{\rm The Feynman diagrams for  the gauge boson  pole and the contact terms  of the  scattering of one $C^{ij}$, one B-field and one graviton on the world volume of D$_3$-brane.}}

The contact terms arise from the first term in the second line of \reef{first113} and from the first term in the second line of \reef{first111}. The massless pole, on the other hand, arises from the second term in the second line of \reef{first113} and from the coupling $Bf$ in the DBI action \reef{DBI}. The amplitude corresponding to the sum of these two Feynman diagrams is the following:
\beqa
{\cal A}^{\rm field}&=&-i\left(\frac{(\pi\alpha')^2T_3}{6}\right)\bigg(\frac{p_2\inn D\inn p_2}{p_3\inn D\inn p_3}p_2^{a_0}p_2^j(\veps_2^S)^{a_0i}(p_3\inn V\inn H_3)^{a_2a_3} \bigg)\eps_{a_0\cdots a_3} (\veps_1)_{ij} \labell{a1}
\eeqa
Similar amplitude as above has been found in \cite{Garousi:2010bm} for the RR $(p-3)$form with one transverse scalar.

Now to compare the above field theory amplitude with the amplitude \reef{totalff}, we have to $\alpha'$-expand the functions $\cI_1, \cI_2, \cI_4$. One may   expand these functions for arbitrary Mandelstam variables using the expansions found in \cite{Becker:2011ar}. However, to simplify the discussion, we use the fact that neither the field theory nor the string theory has  pole $1/p_2\inn p_3$ or $1/p_2\inn D\inn p_3$. So  we expand these functions   for the case that $p_2\inn p_3=p_2\inn D\inn p_3=0$. This expansion  has been found in  \cite{Garousi:2010bm}
\beqa
\cI_1&\!\!\!\!\!=\!\!\!\!\!&-\frac{\pi}{2}\left(\frac{1}{p_1\inn p_2 p_1\inn p_3}-\frac{\pi^2}{6}\bigg[\frac{p_2\inn D\inn p_2}{p_1\inn p_3}+\frac{p_3\inn D\inn p_3}{p_1\inn p_2}\bigg]+\cdots \right)\nonumber\\
\cI_2&\!\!\!\!\!=\!\!\!\!\!&\pi\left(\frac{1}{p_1\inn p_2(p_1\inn p_2+p_1\inn p_3)}-\frac{\pi^2}{6}\frac{p_3\inn D\inn p_3}{p_1\inn p_2}+\cdots\right)\labell{i1i2} \\
\cI_4&\!\!\!\!\!=\!\!\!\!\!&-\pi \left(\frac{1}{2p_1\inn p_2 p_1\inn p_3}+\frac{1}{p_1\inn p_2\,p_3\inn D\inn p_3}-\frac{\pi^2}{12}\bigg[\frac{2p_1\inn N\inn p_3}{p_1\inn p_2}+\frac{2p_2\inn D\inn p_2}{p_3\inn D\inn p_3}+\frac{p_2\inn D\inn p_2}{p_1\inn p_3}\bigg]+\cdots\right)\nonumber 
\eeqa
Note that $\cI_3$ and $\cI_7$ are the same as $\cI_2$ and $\cI_4$, respectively, in which the momentum labels 2,3 are interchanged. The above expansion shows that only $\cI_4$ and $\cI_7$ have massless open string poles. One can easily verify that the field theory amplitude \reef{a1} is the open string pole of $\cI_4$ in the last term in the third line of \reef{totalff}.

The Feynman diagrams in class $(ii)$ are given in figure 2. 
\begin{center} 
\begin{picture}(140,120)(0,0)
\SetColor{Black}
\Line(10,40)(10,90)
\Line(10,90)(30,110)
\Line(30,110)(30,60)
\Line(30,60)(10,40)
\Text(0,108)[]{$D_3$-brane}
%\Text(25,20)[]{(1.b)}
\Vertex(20,90){1.5}\Vertex(20,60){1.5}
\Photon(20,60)(20,90){1}{8} \Text(14,75)[]{$\Phi$}
\Gluon(20,90)(65,90){2}{10} \Text(44,102.5)[]{$h_3$}
\Gluon(20,60)(65,60){3}{4}\Text(42.5,72.5)[]{$B_2$}
\Photon(20,60)(50,35){3}{4} \Text(60,45)[]{$C_1^{(ij)}$}
\end{picture}\begin{picture}(150,130)(0,0)
\SetColor{Black}
\Line(10,40)(10,90)
\Line(10,90)(30,110)
\Line(30,110)(30,60)
\Line(30,60)(10,40)
\Text(0,108)[]{$D_3$-brane}
\Vertex(20,75){1.5}\Vertex(67,75){1.5}
\Gluon(65,75)(100,100){2}{8} \Text(84,101)[]{$h_2$}
\Photon(20,75)(65,75){3}{4} \Text(52,85)[]{$C^{(2)}$}
\Gluon(20,75)(65,39){4}{6}\Text(39,47)[]{$B_3$}
\Photon(67,75)(100,50){3}{4} \Text(83,49)[]{$C_1^{(ij)}$}
%\Text(20,20)[]{(2.a)}
\end{picture}
%\\{\sl Figure 1 :{\rm Feynman diagram for massless open string pole.}}
%\end{center}
%\begin{center}  
\begin{picture}(100,0)(0,0)
\SetColor{Black}
\Line(10,40)(10,90)
\Line(10,90)(30,110)
\Line(30,110)(30,60)
\Line(30,60)(10,40)
\Text(0,108)[]{$D_3$-brane}
%\Text(25,20)[]{(2.b)}
\Vertex(20,75){1.5}\Vertex(67,75){1.5}
\Photon(20,75)(65,75){3}{4} \Text(52,85)[]{$C^{(4)}$}
\Gluon(20,75)(62,43){2}{10}\Text(40,50)[]{$h_2$}
\Gluon(65,75)(100,100){4}{4} \Text(78,100)[]{$B_3$}
\Photon(67,75)(100,50){3}{4} \Text(83,49)[]{$C_1^{(ij)}$}
\end{picture}\end{center}
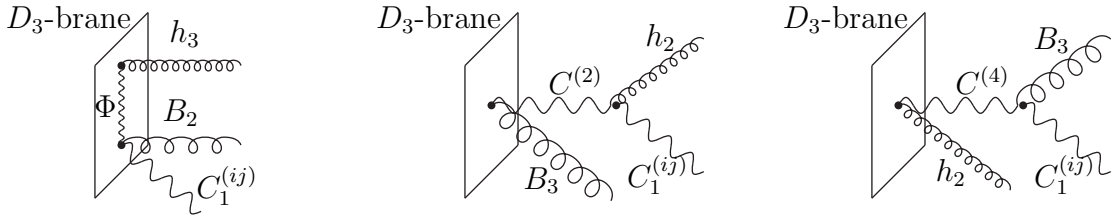
{\sl Figure 2 :{\rm The Feynman diagrams for  the open string transverse scalars  pole and the closed string poles   of the  scattering of one $C^{ij}$, one B-field and one graviton on the world volume of D$_3$-brane.}}

One vertex in the open string amplitude arises from the pull-back and the Taylor expansion  of the linear graviton in the DBI action, and the other one arises from the pull-back of the first term in \reef{first111}. In the closed string poles, the vertices on the brane arise from the action \reef{first111} and the vertices in the bulk arise from the supergravity action \reef{IIB}. We have calculated these amplitudes explicitly. Using the following identity:
\beqa
p_3^{a_2}H_3^{aa_0a_1}\eps_{a_0\cdots a_3}&=&\frac{1}{3}p_3^aH_3^{a_0a_1a_2}\eps_{a_0\cdots a_3}
\eeqa
we have  found exact agreement between the field theory amplitude and the  open and closed string poles at order $\alpha'^2$ which   result from replacing the expansion \reef{i1i2} in the amplitude \reef{totalff}.

\subsection{$C^{(p+1)}$ case}

In this case the assumption that the RR potential has no world volume indices fixes $p=2$. The amplitude is nonzero when the two NSNS states are symmetric or antisymmetric. When the two NSNS states are antisymmetric, the Feynman diagrams in class $(i)$ are given in the figure 3.  
\begin{center} \begin{picture}(150,115)(0,0)
\SetColor{Black}
\Line(10,40)(10,90)
\Line(10,90)(30,110)
\Line(30,110)(30,60)
\Line(30,60)(10,40)
\Text(0,108)[]{$D_2$-brane}
\Vertex(20,90){1.5}\Vertex(20,60){1.5}
\Photon(20,60)(20,90){3}{8} \Text(12,75)[]{$A$}
\Gluon(20,90)(65,90){3}{4} \Text(42.5,102.5)[]{$B_3$}
\Gluon(20,60)(65,60){3}{4}\Text(42.5,70)[]{$B_2$}
\Photon(20,60)(50,35){3}{4} \Text(60,45)[]{$C_1^{(ijk)}$}
%\Text(20,20)[]{(3.a)}
\end{picture}
%\\{\sl Figure 1 :{\rm Feynman diagram for massless open string pole.}}
%\end{center}
%\begin{center}
\begin{picture}(50,0)(0,0)
\SetColor{Black}
\Line(10,40)(10,90)
\Line(10,90)(30,110)
\Line(30,110)(30,60)
\Line(30,60)(10,40)
\Text(0,108)[]{$D_2$-brane}
%\Text(25,20)[]{(1.b)}
\Vertex(20,75){1.5}%\Vertex(20,60){1.5}
%\Photon(20,60)(20,90){1}{8} \Text(14,75)[]{$\Phi$}
\Gluon(20,75)(65,74){3}{5} \Text(63,83)[]{$B_2$}
\Gluon(20,75)(65,97.5){3}{5}\Text(63,106)[]{$B_3$}
\Photon(20,75)(55,50){3}{5} \Text(66.5,56)[]{$C_1^{(ijk)}$}
\end{picture}
%\begin{picture}(50,0)(0,0)
%\SetColor{Black}
%\Line(10,40)(10,90)
%\Line(10,90)(30,110)
%\Line(30,110)(30,60)
%\Line(30,60)(10,40)
%\Text(0,108)[]{$D_2$-brane}
%\Text(25,20)[]{(3.b)}
%\Vertex(20,90){1.5}\Vertex(20,60){1.5}
%\Photon(20,60)(20,90){1}{8} \Text(14,75)[]{$\Phi$}
%\Gluon(20,90)(65,90){2}{10} \Text(44,102.5)[]{$h_3$}
%\Gluon(20,60)(65,60){2}{10}\Text(42.5,72.5)[]{$h_2$}
%\Photon(20,60)(50,35){3}{4} \Text(60,45)[]{$C_1^{(ijk)}$}
%\end{picture}
\end{center}
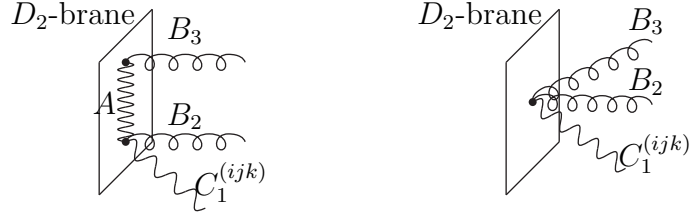
{\sl Figure 3 :{\rm The Feynman diagrams for  the gauge boson  pole and the contact terms  of the  scattering amplitude of one $C^{ijk}$ and two  B-fields   on the world volume of D$_2$-brane.}}

The contact terms arise from the first term in the third line of \reef{first113} and the first term in the third line of \reef{first111}. The open string pole is produced by the coupling in the second term in the second line of \reef{first113}, and by $Bf$ in the DBI action \reef{DBI}. The amplitude corresponding to the sum of these two Feynman diagrams is the following:
\beqa
{\cal A}^{\rm field}&=&i\left(\frac{(\pi\alpha')^2T_2}{6}\right)\bigg(\frac{p_2\inn D\inn p_2}{p_3\inn D\inn p_3}p_2^{a_0}H_2^{ijk}(p_3\inn V\inn H_3)^{a_1a_2} \bigg)\eps_{a_0a_1 a_2}(\veps_1)_{ijk} \nonumber
\eeqa
which is exactly the open string pole in the $\cI_4$ term in \reef{totalfffA}.

The Feynman diagrams in class $(ii)$  are given in the figure 4. \\

 \begin{center} 
%\begin{picture}(100,0)(0,0)
%\SetColor{Black}
%\Line(10,40)(10,90)
%\Line(10,90)(30,110)
%\Line(30,110)(30,60)
%\Line(30,60)(10,40)
%\Text(10,114)[]{$D_2$-brane}
%\Text(25,20)[]{(3.b)}
%\Vertex(20,90){1.5}\Vertex(20,60){1.5}
%\Photon(20,60)(20,90){1}{8} \Text(14,75)[]{$\Phi$}
%\Gluon(20,90)(65,90){2}{10} \Text(44,102.5)[]{$h_3$}
%\Gluon(20,60)(65,60){2}{10}\Text(42.5,72.5)[]{$h_2$}
%\Photon(20,60)(50,35){3}{4} \Text(60,45)[]{$C_1^{(ijk)}$}
%\end{picture}
%\begin{picture}(110,0)(0,0)
%\SetColor{Black}
%\Line(10,40)(10,90)
%\Line(10,90)(30,110)
%\Line(30,110)(30,60)
%\Line(30,60)(10,40)
%\Text(10,114)[]{$D_2$-brane}
%\Vertex(20,75){1.5}\Vertex(67,75){1.5}
%\Gluon(65,75)(100,100){2}{8} \Text(84,101)[]{$h_2$}
%\Photon(20,75)(65,75){3}{4} \Text(52,85)[]{$C^{(3)}$}
%\Gluon(20,75)(65,39){2}{9}\Text(42,47)[]{$h_3$}
%\Photon(67,75)(100,50){3}{4} \Text(80,48)[]{$C_1^{(ijk)}$}
%\Text(20,20)[]{(4.a)}
%\end{picture}
\begin{picture}(150,100)(0,0)
\SetColor{Black}
\Line(10,40)(10,90)
\Line(10,90)(30,110)
\Line(30,110)(30,60)
\Line(30,60)(10,40)
\Text(0,108)[]{$D_2$-brane}
%\Text(25,20)[]{(4.b)}
\Vertex(20,75){1.5}\Vertex(67,75){1.5}
\Photon(20,75)(65,75){3}{4} \Text(52,85)[]{$C^{(1)}$}
\Gluon(20,75)(62,43){4}{4}\Text(38,50)[]{$B_3$}
\Gluon(65,75)(100,100){4}{4} \Text(78,100)[]{$B_2$}
\Photon(67,75)(100,50){3}{4} \Text(81,49)[]{$C_1^{(ijk)}$}
\end{picture}
\begin{picture}(50,0)(0,0)
\SetColor{Black}
\Line(10,40)(10,90)
\Line(10,90)(30,110)
\Line(30,110)(30,60)
\Line(30,60)(10,40)
\Text(0,108)[]{$D_2$-brane}
%\Text(25,20)[]{(4.c)}
\Vertex(20,75){1.5}\Vertex(67,75){1.5}
\Photon(20,75)(65,75){3}{4} \Text(52,85)[]{$C^{(5)}$}
\Gluon(20,75)(62,43){4}{4}\Text(39,50)[]{$B_3$}
\Gluon(65,75)(100,100){4}{4} \Text(78,100)[]{$B_2$}
\Photon(67,75)(100,50){3}{4} \Text(81,49)[]{$C_1^{(ijk)}$}
\end{picture}
\end{center}
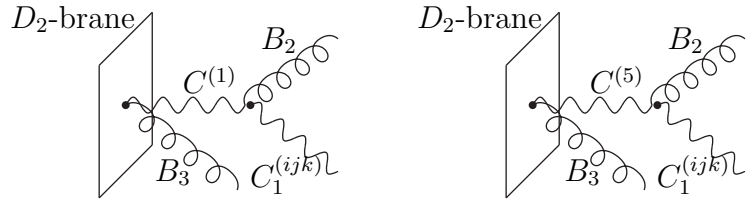
{\sl Figure 4 :{\rm The Feynman diagrams for    closed string poles   of the  scattering amplitude of one $C^{ijk}$ and two B-fields   on the world volume of D$_2$-brane.}}
 
There is no $B\Phi$ coupling in the D-branes action, so there is no open string pole. This is consistent with the amplitude \reef{totalfffA}. For the closed string poles, the vertices on the brane can be read from the couplings \reef{first111} and the vertices in the bulk can be read from the supergravity action \reef{IIA}. We have computed these amplitudes and found exact agreement with the massless closed string poles in \reef{totalfffA}.

When the two NSNS states are symmetric, the actions \reef{first111} and \reef{first113} produce  no  Feynman diagrams in the class $(i)$ which is consistent with the amplitude \reef{totalfffS}. The massless open string pole resulting from the function $\cI_4$ in \reef{totalfffS} is in fact the transverse scalar pole which is in the   the Feynman diagrams in class $(ii)$. The Feynman diagrams in the class $(ii)$ in this case  are given in the figure 5. \\
 
\begin{center} 
\begin{picture}(150,100)(0,0)
\SetColor{Black}
\Line(10,40)(10,90)
\Line(10,90)(30,110)
\Line(30,110)(30,60)
\Line(30,60)(10,40)
\Text(0,108)[]{$D_2$-brane}
%\Text(25,20)[]{(3.b)}
\Vertex(20,90){1.5}\Vertex(20,60){1.5}
\Photon(20,60)(20,90){1}{8} \Text(14,75)[]{$\Phi$}
\Gluon(20,90)(65,90){2}{10} \Text(44,102.5)[]{$h_3$}
\Gluon(20,60)(65,60){2}{10}\Text(42.5,72.5)[]{$h_2$}
\Photon(20,60)(50,35){3}{4} \Text(60,45)[]{$C_1^{(ijk)}$}
\end{picture}
\begin{picture}(50,0)(0,0)
\SetColor{Black}
\Line(10,40)(10,90)
\Line(10,90)(30,110)
\Line(30,110)(30,60)
\Line(30,60)(10,40)
\Text(0,108)[]{$D_2$-brane}
\Vertex(20,75){1.5}\Vertex(67,75){1.5}
\Gluon(65,75)(100,100){2}{8} \Text(84,101)[]{$h_2$}
\Photon(20,75)(65,75){3}{4} \Text(52,85)[]{$C^{(3)}$}
\Gluon(20,75)(65,39){2}{9}\Text(42,47)[]{$h_3$}
\Photon(67,75)(100,50){3}{4} \Text(80,48)[]{$C_1^{(ijk)}$}
%\Text(20,20)[]{(4.a)}
\end{picture}
\end{center} 
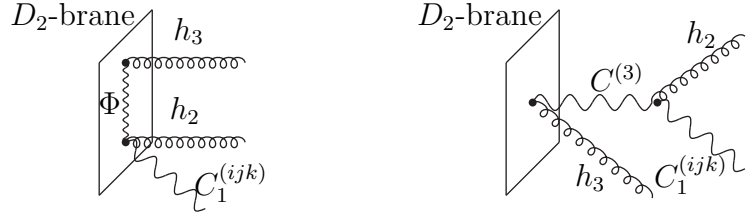
{\sl Figure 5 :{\rm The Feynman diagrams for   transverse scalars pole and for the  closed string pole  of the  amplitude of one $C^{ijk}$ and two gravitons   on the world volume of D$_2$-brane.}}

The open string pole arises from the pull-back of the first term in the second line of \reef{first111}, and  from the pull-back and the Taylor expansion  of the linear graviton in the DBI action. The closed string pole arise from the brane couplings in the second line of \reef{first111} and the bulk couplings in \reef{IIA}.  Here again we have   found exact agreement with the string theory results in \reef{totalfffS}.

 \subsection{$C^{(p+3)}$ case}
 
In this case for $p=1$ there are no world volume indices carried by the RR potential. The scattering amplitude in field theory is nonzero when one of the NSNS states is symmetric and the other one is antisymmetric. The D-brane action at order $\alpha'^2$ produces  no Feynman diagram in class $(i)$ which is consistent with the string theory amplitude \reef{totalffff}. The Feynman diagrams in the class $(ii)$  are shown in 
  figure 6.
\begin{center} \begin{picture}(155,130)(0,0)
\SetColor{Black}
\Line(10,40)(10,90)
\Line(10,90)(30,110)
\Line(30,110)(30,60)
\Line(30,60)(10,40)
\Text(0,108)[]{$D_1$-brane}
%\Text(25,20)[]{(5.a)}
\Vertex(20,90){1.5}\Vertex(20,60){1.5}
\Photon(20,60)(20,90){1}{8} \Text(14,75)[]{$\Phi$}
\Gluon(20,90)(65,90){2}{10} \Text(44,102.5)[]{$h_2$}
\Gluon(20,60)(65,60){3}{4}\Text(42.5,72.5)[]{$B_3$}
\Photon(20,60)(50,35){3}{4} \Text(60,45)[]{$C_1^{(ijkl)}$}
\end{picture}
\begin{picture}(150,0)(0,0)
\SetColor{Black}
\Line(10,40)(10,90)
\Line(10,90)(30,110)
\Line(30,110)(30,60)
\Line(30,60)(10,40)
\Text(0,108)[]{$D_1$-brane}
%\Text(25,20)[]{(5.b)}
\Vertex(20,75){1.5}\Vertex(67,75){1.5}
\Photon(20,75)(65,75){3}{4} \Text(52,85)[]{$C^{(4)}$}
\Gluon(20,75)(62,43){4}{4}\Text(40,47)[]{$B_3$}
\Gluon(65,75)(100,100){2}{8} \Text(78,100)[]{$h_2$}
\Photon(67,75)(100,50){3}{4} \Text(81,47)[]{$C_1^{(ijkl)}$}
\end{picture}
\begin{picture}(100,0)(0,0)
\SetColor{Black}
\Line(10,40)(10,90)
\Line(10,90)(30,110)
\Line(30,110)(30,60)
\Line(30,60)(10,40)
\Text(0,108)[]{$D_1$-brane}
%\Text(25,20)[]{(5.c)}
\Vertex(20,75){1.5}\Vertex(67,75){1.5}
\Photon(20,75)(65,75){3}{4} \Text(52,85)[]{$C^{(2)}$}
\Gluon(20,75)(62,43){2}{9}\Text(44,46)[]{$h_2$}
\Gluon(65,75)(100,100){4}{4} \Text(78,100)[]{$B_3$}
\Photon(67,75)(100,50){3}{4} \Text(81,46)[]{$C_1^{(ijkl)}$}
\end{picture}
\end{center}
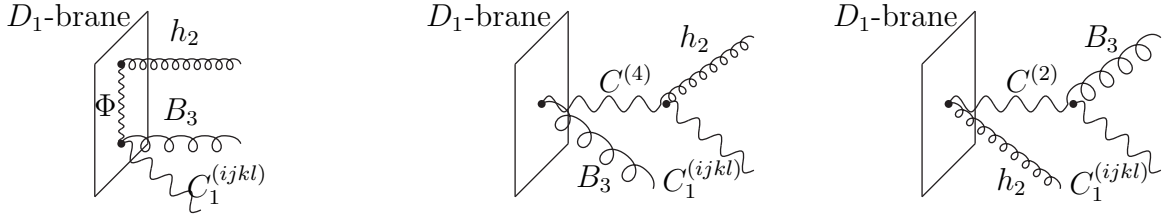
{\sl Figure 6 :{\rm The Feynman diagrams for   transverse scalars pole and for the  closed string poles  of the  amplitude of one $C^{ijkl}$, one B-field  and one graviton.}}

The explicit calculation again reproduces the massless poles in \reef{totalffff} at order $\alpha'^2$.

\subsection{$C^{(p+5)}$  case}

In this case the assumption that the RR potential has no world volume indices fixes $p=0$. The field theory amplitude is nonzero when the two NSNS states are antisymmetric. There is only one Feynman diagram which is given in   figure 7.
\begin{center}
\begin{picture}(130,115)(0,0)
\SetColor{Black}
\Line(10,40)(10,90)
\Line(10,90)(30,110)
\Line(30,110)(30,60)
\Line(30,60)(10,40)
\Text(0,108)[]{$D_0$-brane}
\Vertex(20,75){1.5}\Vertex(67,75){1.5}
\Photon(20,75)(65,75){3}{4} \Text(52,85)[]{$C^{(3)}$}
\Gluon(20,75)(62,43){4}{4}\Text(48,39)[]{$B_3$}
\Gluon(65,75)(100,100){4}{4} \Text(78,100)[]{$B_2$}
\Photon(67,75)(100,50){3}{4} \Text(91,41)[]{$C_1^{(ijklm)}$}
\end{picture}\end{center}
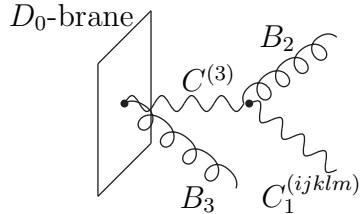
{\sl Figure 7 :{\rm The Feynman diagram for     the  closed string pole  of the scattering  amplitude of one $C^{ijklm}$ and two   B-fields on the world volume of D$_0$-brane.}
}

Using the following identity:
\beqa
\veps_{a_0}p_2^{a_0}(p_3\inn V\inn H_2)^{ij}=\veps_{a_0}p_3^{a_0}(p_2\inn V\inn H_2)^{ij}
\eeqa
which is easy to verify using the fact that there is only one world volume direction, one finds the field theory result to agree  exactly with the string theory amplitude \reef{fffff} at order $\alpha'^2$. 
This ends our illustration of the precise consistency between the T-dual multiplet $\textbf{A}$ \reef{AA}  and the field theory couplings \reef{first111}, \reef{first113} and the supergravity couplings. 
 
 \section{Discussion}
 
 In this paper, we  have shown that the T-dual Ward identity and the Ward identity of the gauge symmetries can be used to find new S-matrix elements from a given S-matrix element. We have shown that the T-duality, in general, can not capture the new S-matrix elements fully. The T-dual multiplet \reef{mul} corresponding to the S-matrix element of one RR $(p-3)$-form which has two transverse indices, and two NSNS states    is fully captured by the T-duality. However, the  multiplet   \reef{TA} corresponding to the S-matrix element of one RR $(p-3)$-form which has one transverse index, and two NSNS states   contains the S-matrix elements which do not satisfy the Ward identity of the NSNS gauge transformations. This indicates there is another T-dual multiplet   \reef{TA'} which must be added to \reef{TA} to fully produce the new S-matrix elements. 
 
  The consistency of the S-matrix elements with the linear T-duality that we have used in this paper does not fix the terms in the S-matrix elements which have momentum along the Killing direction $y$. When $y$ is a world volume coordinate in the original theory, then in the T-dual theory it becomes transverse coordinate. So our prescription for finding the T-dual multiplets outlined in section 2 can not capture the terms of the S-matrix elements which  have an extra factor of $p_1^k$, $p_2^k$ or $p_3^k$  relative to the original S-matrix element on which the consistency with the linear T-duality is applied.  As a result, the T-dual Ward identity can not capture the remnant  multiplets     that are  proportional to $p_1^k$, $p_2^k$ or $p_3^k$. The remnant terms proportional to $p_1^k$ make the amplitude to satisfy the Ward identity of the RR gauge transformations in which we have not been interested in this paper. Whereases the remnant terms proportional to $p_2^k$ or $p_3^k$ make the amplitude to satisfy the Ward identity of the NSNS gauge transformations. This explain   why the multiplet \reef{mul} does not have the remnant terms proportional to $p_2^k$ or $p_3^k$. This T-dual multiplet is proportional to $p_2^ip_3^j$, hence, the  remnant T-dual multiplet   is proportional to $p_2^ip_3^jp_3^k$ or $p_2^ip_3^jp_2^k$. However, this multiplet  is zero because these momenta contract with the RR potential which is totally antisymmetric. On the other hand the multiplet $\cA$ in \reef{TA} is proportional to $p_2^i$ or $p_3^i$. The remnant T-dual multiplet is then proportional to $p_2^ip_3^j$ which produce non-vanishing terms when contracted with the RR potential. 
 
 The above discussion can be applied for the S-matrix element of one RR $(p-3)$-form which has no transverse indices, and two NSNS states. Since the RR $(p-3)$-form has no transverse index, this S-matrix element has no momentum $p_2^i$ or $p_3^i$ contracted with the RR potential, so its corresponding T-dual multiplet does not have such factor.  In this case there are two remnant  T-dual multiplets that are intertwined  by the NSNS gauge transformations. One of them is proportional to $p_2^i$ or $p_3^i$, and the other one is proportional to $p_2^ip_3^j$. The components of multiplets are connected as the following:
\beqa
\matrix{\cA_0& \!\!\!\!\!\rightarrow \!\!\!\!\!& \cA_1& \!\!\!\!\!\rightarrow \!\!\!\!\!&\cA_2&  \!\!\!\!\!\rightarrow \!\!\!\!\! & \cA_3& \!\!\!\!\!\rightarrow \!\!\!\!\!&\cA_4\cr 
&&\downarrow & &\downarrow&& \downarrow&& \downarrow\cr
&&  \cA'_1& \!\!\!\!\!\rightarrow \!\!\!\!\!&\cA'_2& \!\!\!\!\!\rightarrow \!\!\!\!\!&\cA'_3& \!\!\!\!\!\rightarrow \!\!\!\!\!&\cA'_4\cr
&&&& \downarrow & &\downarrow&& \downarrow \cr
&&&&\cA''_2& \!\!\!\!\!\rightarrow \!\!\!\!\!&\cA''_3& \!\!\!\!\!\rightarrow \!\!\!\!\!&\cA''_4 }
   \eeqa  
 The $\cA_0$ component of the above multiplets has been calculated in \cite{Garousi:2011ut,Becker:2011ar}. It would be interesting to find all other components by the T-dual   and the gauge transformation Ward identities, and find their corresponding D-brane couplings at order $\alpha'^2$ \cite{BG}.

One may use  these Ward identities to find the connections between the S-matrix elements of one RR $n$-form   in the $(-3/2,-1/2)$-picture  and $l$ massless NSNS vertex operators in $(0,0)$-picture on the world volume of a D$_p$-brane. The amplitudes are again proportional to the trace factor \reef{trace} in which $m=0,2,\cdots, 4l$. The totally antisymmetric factor $A_{[\alpha_1\cdots \alpha_m]}$ for $m=4l$ has  the momentum factor  $p_2p_2p_3p_3\cdots p_{l+1}p_{l+1}$. So the trace factor for $m=4l$ is non-zero when the RR potential has at least  $l$ transverse indices which contract with $p_2p_3\cdots p_{l+1}$. Then for $n=p+1-2l$, the trace is non-zero when all other indices of the RR potential contract with the world volume form. The  T-dual multiplet corresponding to this term is  
 \beqa
 A_l\rightarrow A_{l+1}\rightarrow \cdots\rightarrow A_{l+2^l}
 \eeqa 
 where $2^l$ is the number of the indices of the NSNS polarization tensors in $A_l$ which carry the world volume indices. In the last component $A_{l+2^l}$ all indices of the NSNS polarization tensors are transverse indices. In this case there is no remnant T-dual multiplet because the remnant multiplet would have an extra factor of  the momentum of one of the NSNS states which contract with the RR potential.  When the RR $(p+1-2l)$-form has $l-1$ transverse indices, there are two T-dual multiplets    that are intertwined by the gauge invariance, when the RR $(p+1-2l)$-form has $l-2$ transverse indices, there are three T-dual multiplets    that are intertwined by the gauge invariance, and so on.
 
The first components of the T-dual multiplets that we have studied in this paper are in terms of the RR potential with  specific indices. This makes all the other components to be in terms of the appropriate RR potentials. The multiplets are proportional to the RR momenta which is a necessary condition for satisfying the Ward identity associated with the RR gauge transformations. The multiplet \reef{mul} can be combined with some terms of the multiplet $\cA$ to become invariant under the RR gauge transformations. All other terms in the multiplet $\textbf{A}$ \reef{AA} would be combined with the T-dual multiplets corresponding to the RR $(p-3)$-form with no transverse index to become invariant.     Alternatively, one may rewrite the first component of a multiplet in terms of the RR field strength, \ie      the combination of the last terms in the second, the third and the fourth lines of  \reef{n=1polarization}, and the amplitude \reef{amp5}    can be written in terms of  the RR field strength $(F_1^{(p-2)})_{ij}{}^{a_5\cdots a_p}$. The other non-zero amplitudes for the RR field strengths   $(F_1^{(p-2)})_{i}{}^{a_4\cdots a_p}$ and $(F_1^{(p-2)}) ^{a_3\cdots a_p}$  have been found in \cite{Garousi:2011ut}.   Then one imposes the method we have advertised in this paper to find all other components in terms of the RR field strength. This latter method has the advantage that the Ward identity corresponding to the RR gauge transformations satisfies automatically. In particular, this makes the calculations toward finding the remnant T-dual multiplets to be easier.  
 
 We have found  connections between   S-matrix  elements by  imposing the T-dual Ward identity on a given S-matrix element. One may find other connections by imposing the S-dual Ward identity \cite{Garousi:2011vs,Garousi:2011jh, Garousi:2012gh}. Since the D$_3$-brane is invariant under the S-duality, the S-matrix elements on its world volume can be written in manifestly S-dual invariant form. In this way, one can find new connections between some S-matrix elements. For example consider the amplitude \reef{A''1} for $p=3$. Using the fact that the graviton and the RR four-form are invariant under the S-duality, one finds that the last terms in this amplitude is invariant. However, the $B$-field and the RR two-form transform as doublet under the S-duality. As a result, the first term in this amplitude can be extended to invariant form by adding to it the following S-matrix element of three RR states: 
 \beqa
{ A}_4(C^{(4})\sim T_3 e^{2\phi_0}(\veps_1^{(4)})_{ijkl} \eps_{a_0\cdots a_3}p_2^{a_0}p_3^{a_1}p_2^ip_3^j  (\veps_2^{(2)})^{kl}(\veps_3^{(2)})^{a_2a_3}  
 \cI_1\delta^{p+1}(p_1^a+p_2^a+p_3^a)+(2\leftrightarrow 3)\nonumber
\eeqa
 where now the tensors $\veps_2^{(2)}$ and $\veps_3^{(2)}$ are the RR two-form polarizations. One may then use the T-dual ward identity to find the T-dual multiplet corresponding to the above amplitude. It would be interesting to find such S-matrix elements which are connected to the T-dual multiplets that we have found in this paper and compare the results  with explicit calculations.

{\bf Acknowledgments}:   This work is supported by Ferdowsi University of Mashhad under grant 2/21644-1391/02/19.

\end{document}